\documentclass{aa}
\usepackage{graphicx}
\usepackage{natbib}
\bibpunct{(}{)}{;}{a}{}{,}

\def\beq{\protect \begin{equation} \protect}
\def\eeq{\protect \end{equation} \protect}
\def\beqa{\protect \begin{eqnarray}}
\def\eeqa{\protect \end{eqnarray}}
\def\bc{\begin{center}}
\def\ec{\end{center}}
\newcommand{\pacz}{Paczy\'nski}
\newcommand{\Pacz}{Paczy\'nski}

\newcommand{\MC}{Monte Carlo}

\begin{document}

\titlerunning{Evidence for a MACHO contribution to Galactic Halos}

\title{POINT-AGAPE Pixel Lensing Survey of M31 :\\
Evidence for a MACHO contribution to Galactic Halos}

\authorrunning{POINT-AGAPE}
\author{{S.~Calchi Novati \inst{1} \and S.~Paulin-Henriksson \inst{2} \and
J.~An \inst{3} \and P.~Baillon \inst{4} \and V.~Belokurov \inst{3} \and
B.J.~Carr \inst{5} \and M.~Cr{\'e}z{\'e} \inst{2,6} \and N.W.~Evans \inst{3} \and
Y.~Giraud-H{\'e}raud \inst{2} \and A.~Gould \inst{7}
\and P.~Hewett \inst{3} \and Ph.~Jetzer \inst{1} \and J. Kaplan \inst{2} \and E.~Kerins \inst{8} 
\and S.J.~Smartt \inst{3,9} \and C.S.~Stalin \inst{2} \and Y.~Tsapras
\inst{5} \and  M.J.~Weston \inst{5}}
\centerline{(The POINT--AGAPE Collaboration)}}

\institute{
Institute for Theoretical Physics,  University of Z\"urich, Winterthurerstrasse 190,
CH-8057 Z\"urich, Switzerland \and APC\thanks{UMR 7164(CNRS, Universit\'e Paris 7, CEA, Observatoire
  de Paris)}, 11 Place Marcelin Berthelot, F-75231 Paris, France \and
Institute of Astronomy, Madingley Road, Cambridge CB3 0HA, UK \and
CERN, 1211 Gen{\`e}ve, Switzerland \and
Astronomy Unit, School of Mathematical Sciences, Queen Mary,
University of London, Mile End Road, London E1 4NS, UK \and
Universit\'e Bretagne-Sud, campus de Tohannic, BP 573, F-56017
Vannes Cedex, France \and
Department of Astronomy, Ohio State University, 140 West 18th
Avenue, Columbus, OH 43210, US \and
Astrophysics Research Institute, Liverpool John Moores University, Twelve Quays House,
Egerton Wharf, Birkenhead CH41 1LD, UK \and
Department of Pure and Applied Physics, The Queen's University of Belfast, Belfast BT7 1NN, UK
}
\date{Received  26 March 2005/ Accepted 8 August 2005}

\abstract{ The POINT-AGAPE collaboration is carrying out a search
for gravitational microlensing toward M31 to reveal galactic dark
matter in the form of MACHOs (Massive Astrophysical Compact Halo
Objects) in the halos of the Milky Way and M31. A high-threshold
analysis of 3 years of data yields 6 bright, short--duration
microlensing events, which are confronted to a simulation of the
observations and the analysis. The observed signal is much larger
than expected from self lensing alone and we conclude, at the 95\%
confidence level, that at least 20\% of the  halo mass in the
direction of M31  must be in the form of MACHOs if their average
mass lies in the range   0.5-1 M$_\odot$. This lower bound drops
to 8\% for MACHOs with masses $\sim 0.01$ M$_\odot$. In addition,
we discuss a likely binary microlensing candidate  with caustic
crossing. Its location, some 32' away from the centre of M31,
supports our conclusion that we are detecting a MACHO signal in
the direction of M31.  } \maketitle

\section{Introduction}

Gravitational microlensing, as first noted by \citet{pacz86},  is a powerful
tool for the detection of massive astrophysical halo compact objects (MACHOs),
a possible component of dark matter halos.
Observations toward the Magellanic Clouds by the first
generation of microlensing surveys yielded important constraints on
the Milky Way (MW) halo. The EROS
collaboration obtained an upper limit ($f<20\%$) on the contribution by MACHOs to
a standard MW halo \citep{eros03}, and the results
of their latest analysis strengthen this conclusion \citep{eros05}.
Also, according to the MACHO
collaboration \citep{macho00}, the optical depth toward the Large Magellanic Cloud is too
large by a factor $\sim 5$ to be accounted for by known populations of
stars. Indeed, further analysis recently confirmed these results \citep{bennett05a,bennett05b}.
This excess is attributed to MACHOs of
mass $\sim 0.4$ M$_\odot$ in the MW halo
contributing $f\sim 20\%$, although this result has been
challenged by several authors (e.g \citealt{jetzer02,belokurov04}). These
exciting and somewhat contradictory results challenge us
to probe the MACHO distribution along different MW
lines of sight and in different galaxies.

M31, being both nearby and similar to the MW,
is a suitable target for such a search \citep{crotts92,baillon93}.
It allows us to explore the MW halo along a different line of sight.
It has its own halo that can be studied globally, and its high inclination
is expected to give a strong gradient in the spatial
distribution of microlensing events \citep{crotts92,jetzer94}.
We note, however, that the latter feature, which was
at first believed to provide an unmistakable
signature for M31 microlensing halo events, seems to be shared,
at least to some extent, by the variable star population
within M31 \citep{an04b}.

Several collaborations have undertaken searches for microlensing
toward M31: AGAPE \citep{agape99}, SLOTT-AGAPE \citep{novati03},
MEGA \citep{dejong04}, Columbia-VATT \citep{ugl04},
WeCAPP \citep{riffeser03} and NainiTal \citep{nainital04}.
Up to now, while some microlensing events have been detected,
no firm conclusion about their physical meaning has  been reported.
In particular, the POINT-AGAPE collaboration has presented a first
analysis focused on the search for bright, short--duration
microlensing events \citep{auriere01,paulin03}.

In this paper, we report the first constraints
on the MACHO fraction of the combined MW and M31 halos
along the line of sight to M31.
We give a complete account of our systematic  search
for bright short-duration events, present the 6 selected microlensing
events, and then describe the simulation used
to predict the characteristics  of the expected events and their
frequency. We
proceed in two steps: a \MC\ simulation produces
an initial (quantifiably over-optimistic) estimate
of the number of expected events, then a simulation of
events (hereafter referred to as ``event simulation'') on the actual
images allows us to assess the detection efficiency of
the analysis pipeline for the type of events produced by the \MC.

In the search for a MACHO signal we must deal with two main backgrounds:
(i) variable stars masquerading as microlensing events
and (ii) self-lensing events (for which both the lens and the source are part
of the luminous components of M31 or MW).
We eliminate the first (see below) and partially isolate the
second using their distinctive spatial distribution.

The paper is organised as follows. In Sect. \ref{sec:analysis},
we recall the observational setup and then describe
our analysis pipeline. The detected
microlensing signal is discussed in Sect. \ref{sec:events}.
In Sect. \ref{sec:MC} we describe the \MC\  simulation
of the experiment and describe its predictions. In Sect. \ref{sec:eff},
we evaluate the detection efficiency of the pipeline. In Sect. \ref{sec:results},
we summarise the analysis and discuss what conclusions can be drawn
about the fraction of M31 and MW halos in the form of MACHOs.

\section{Data analysis} \label{sec:analysis}

\subsection{Setup, data acquisition and reduction} \label{sec:setup}

\begin{figure}
\resizebox{\hsize}{!}{\includegraphics{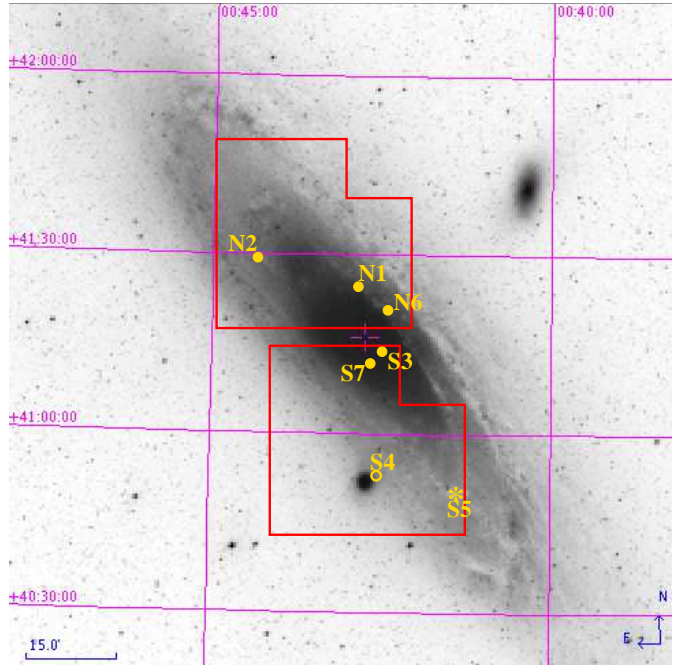}}
\caption{
Projected on M31, we display the boundaries of the observed fields
(red lines), and the centre of M31 (cross).
Circles mark the positions of the 6 microlensing events issued
from the selection pipeline (Sect. \ref{sec:pa-events}).
The open circle (S4) corresponds to the event seen toward M32.
The star (S5) to the binary event candidate
discussed in Sect. \ref{sec:pa-s5}.}
\label{fig:field}
\end{figure}

In this work we analyse data acquired during three seasons of
observation using the Wide Field Camera (WFC) mounted on the 2.5m
Isaac Newton Telescope (INT) \citep{auriere01,an04b}.
A fourth year of data is currently being analysed. Two fields,
each $\sim 0.3$ deg$^2$, north and south of the M31 centre are
monitored (Fig. \ref{fig:field}). The data are taken in two
passbands (\emph{Sloan} $r$ and either \emph{Sloan} $g$ or
\emph{Sloan} $i$), with exposure time between 5 and 10 minutes per
night, field and filter. Each season of observation lasts about
six months, but with very irregular sampling (especially during
the third one). Overall, for $r$ data, we have about 120 nights of
observation. At least two exposures per field and filter were made
each night with a slight dithering. Although they are combined in
the light curve analysis, they allow us to assess, if necessary,
the reality of detected variations by direct inspection of single
images.

Data reduction is performed following \citet{agape97}, \citet{novati02}
and \citet{paulin03}.
Each image is geometrically and photometrically aligned
relative to a reference image (one per CCD, the geometric
reference being the same for all the filters).
Ultimately, in order to deal with seeing variations, we  substitute
for the flux of each pixel that of the corresponding
7-pixel square "superpixel" centred on it, the pixel size
being 0.33'', and we then
apply an empirical correction, again calibrating
each image against the reference image.

\subsection{Analysis: selection of microlensing events} \label{sec:ana-ml}

To search for microlensing events, we use the ``pixel-lensing''
technique \citep{baillon93,gould96,agape97}, in which one monitors the flux variations
of unresolved sources of each pixel element of the image.

A difficulty, specific to pixel lensing, is that genuine
microlensing events might be polluted by one of the numerous
variable objects present in the neighbouring pixels. To avoid
loosing too many \emph{bona fide} microlensing events while accounting for the
variable background, we look for microlensing-like variations even
on those light curves on which a second bump is detected. In
particular, on each light curve we first look for and characterise
mono-bump variations for each season separately. Only as a final step
do we test for bump uniqueness on the complete light curve in a
loose way as explained below. This test allows for the presence of
variable stars within the superpixel containing the lensed source
and so, as a bonus, in principle could allow us to detect
microlensing of variable objects.

In addition to the physical background of variable stars,
the search for microlensing-like flux variations,
in particular the short ones, is plagued by the detection of ``fake'' variations,
mainly due to bad images, defects on the CCD, saturated pixels
associated with extremely bright stars, and
cosmic rays  (these issues are discussed in more detail in
\citealt{london}). The only safe way to remove these artefacts
is to visually inspect the images around the time of maximum,
although there may be other useful hints, such as an anomalous distribution of the times
of maximum or in the spatial distribution. To obtain a ``clean'' set of variations
we first run the complete pipeline, identify and remove bad images,
and mask bad pixels. Then, we rerun the analysis from scratch.

Before proceeding further with the pipeline, we mask
a small region right around the centre of M31,
$\sim 1'\times 1'$, where, in addition to
problems caused by saturation,
the severe uncertainty in modelling the experiment
would prevent us from drawing any significant conclusion
about the physical implication of any result we might obtain.

As a first step, we establish a catalogue of significant flux variations
(using the $r$ band data only, which are both better sampled
and less seriously contaminated
by intrinsically variable stars than the $i$ band data).
Following \citet{novati03}, we use the two estimators,
$L$ and $Q$, which are both monotonic functions
of the significance of a flux variation, to select candidates.
Note that the previous POINT-AGAPE selections presented in
\citet{paulin03,an04b,belokurov05}, have been carried
out using the $L$ estimator only.

We define
\begin{equation} \label{likelihood}
L = -\ln\left(\Pi_{j\in bump}P(\Phi|\Phi>\Phi_{j})\right)
\;\;\mbox{given}\;\; \bar{\Phi}_{bkg},\, \sigma_{j}\,,
\end{equation}
where
\begin{equation}
P(\Phi|\Phi>\Phi_{j}) =  \int_{\Phi_{j}}^\infty d\Phi {\frac{1}
{\sigma_{j}\sqrt {2\pi}}} \exp\left[{-\frac{(\Phi
-\bar{\Phi}_{bkg})^2}{2\sigma_{j}^{2}}}\right];
\end{equation}
$\Phi_j$ and $\sigma_j$ are the flux and associated error
in a superpixel at time $t_j$, $\bar{\Phi}_{bkg}$ is an estimator of the
baseline level, defined as the minimum value of a sliding average
over 18 epochs. A ``bump'' is defined as a positive variation with
at least 3 consecutive points more than $3\,\sigma$ above the
baseline, and it is regarded as ending after two consecutive points
less than this threshold.  We define
\begin{equation}\label{rapp}
  Q\equiv \frac{\chi^2_{const}-\chi^2_{pacz}}{\chi^2_{pacz}/\rm dof}\,,
\end{equation}
where $\chi^2_{const}$ is calculated with respect to the
constant-flux
hypothesis and $\chi^2_{pacz}$ is the $\chi^2$ calculated with respect to a
\Pacz\ fit. Let us stress that $Q$ is evaluated for each full
season, while $L$ is evaluated only inside the bump.
At this point, we keep only light curves with $Q>100$.
Since $Q$ is biased toward mono-bump variations, this step
allows us to remove the unwanted background of short-period variable stars.

Although it has already been described in \citealt{novati02} ,
we return to a crucial
step of the above analysis. For each physical variation, there appears a whole
cluster of pixels with $Q>100$ (with typical size range from 4 to 30
pixels). From the $Q$ values of all light curves, we construct a
$Q$ map for each season. We then proceed to the actual localisation
of the physical variations\footnote{We use here
a software developed within the AGAPE collaboration.}. First we
identify the clusters (which appear as hills on the map). Then we
locate the centre of the cluster as the pixel with the highest value of the
$L$ estimator. The main difficulty arises from the overlap of
clusters. Indeed we must balance the search for faint variations with
the need to separate neighbouring clusters. In the following, we will
refer to this crucial part of the analysis as ``cluster detection''. It must
be emphasised that this step cannot be carried out on separate
light curves, but requires using $Q$ maps. The impossibility of
including this cluster detection in the \MC\  (Sect. \ref{sec:MC})
gives us one of the strongest motivations for
the detection efficiency analysis described in Sect. \ref{sec:eff}.
After the clusterisation, we are left with $\sim 10^5$ variations.

The following part of the analysis
is carried out working only on pixel \emph{light curves}.

As a second cut, we remove flux variations having  too
small a signal-to-noise ratio (most likely due to noise)
by demanding $L_1>40$, $L_1$ being associated with the bump.
If the light curve shows a second bump
over the three seasons, characterised by $L_2$, we then demand that
this satisfies $L_2 < 0.5 L_1$. As we are only looking for bright bumps
(see below), we consider such a significant second bump to indicate
that these bumps most likely  belong to a variable star.

We estimate the probability for the lightcurve of a given event
to be contaminated by a nearby variable source as the fraction
of pixels showing a significant variation, $L_1>40$.
This fraction stronlgy depends on the distance from the centre of M31:
from $\sim 10\%-20\%$ in the inner M31 region,
within an angular radius of $8'$, down to $\sim 8\%$ in the outer region.

We characterise the shape of the variation by studying its
compatibility with a \citet{pacz86} shape. We perform a two-band
7-parameter fit: the Einstein time, $t_{\textrm E}$, the impact
parameter, $u_0$, the time at maximum magnification $t_0$, and the
band dependent flux of the unresolved source, $\phi^*$, and the
background flux, $\phi_b$, of the bump in each of 2 bands ($r$ and
either $i$ or $g$ according to the available data along the
bump)\footnote{Note that, even if it does not contain any
astrophysical information, we must include the background pixel
flux as a parameter in the fit to take into account its
statistical fluctuation when we estimate the parameters of the
\Pacz\ curve.}. Throughout the analysis we use, as an
observable time width, the full-width-half-maximum (FWHM) in time
of the \pacz\ curve, $t_{1/2}$, and the flux increase $\Delta\Phi$
of the bump, both of which are functions of the degenerate
parameters $t_{\textrm E},\, u_0$ and $\phi^*$ \citep{gould96}.
Using the flux deviation in the two bands, we evaluate in the
standard Johnson/Cousins magnitude system $R(\Delta\Phi)$, the
``magnitude at maximum'' of the bump, and its colour, either $V-R$
or $R-I$. The simultaneous \pacz\ fit in two bands effectively
provides a test of the achromaticity expected for microlensing
events.

As a third cut, we use the goodness of the \pacz\ fit
as measured by the reduced $\chi^2$. For short events, the behaviour of the
baseline would dominate the $\chi^2$. To avoid this bias, we perform
the fit in a smaller ``bump region'' defined as follows. A first
\pacz\ fit on the whole baseline provides us with the value of the
baseline flux $\phi_b$ and first estimates of the time of maximum
magnification $t_0$ and the time width $t_{1/2}$. Using
these values we compare two possible definitions of the bump region and use
whichever is the larger of: (i) the time interval
inside $t_0\pm 3\,t_{1/2}$, and (ii) the time interval that begins
and ends with the first two consecutive points less than 3 $\sigma$ above the
background on both sides of $t_0$. The final \pacz\ fit is carried
out in this ``bump region'' with the basis flux $\phi_b$ fixed in
both colours, and this fit provides the values of the 5 remaining
parameters.

Our third selection criterion excludes light curves with
$\chi^2/\textrm{dof}>10$.

We fix this threshold high enough
to accept light curves whose shapes slightly deviate
from the \pacz\ form, either because of a real deviation in the
microlensing signal, as is the case for the microlensing event
PA-99-N2 discussed by \citet{an04a}, or because the signal may be
disturbed by artefacts or by some nearby variable stars.

\begin{figure}
\begin{center}
{\includegraphics[scale=0.33]{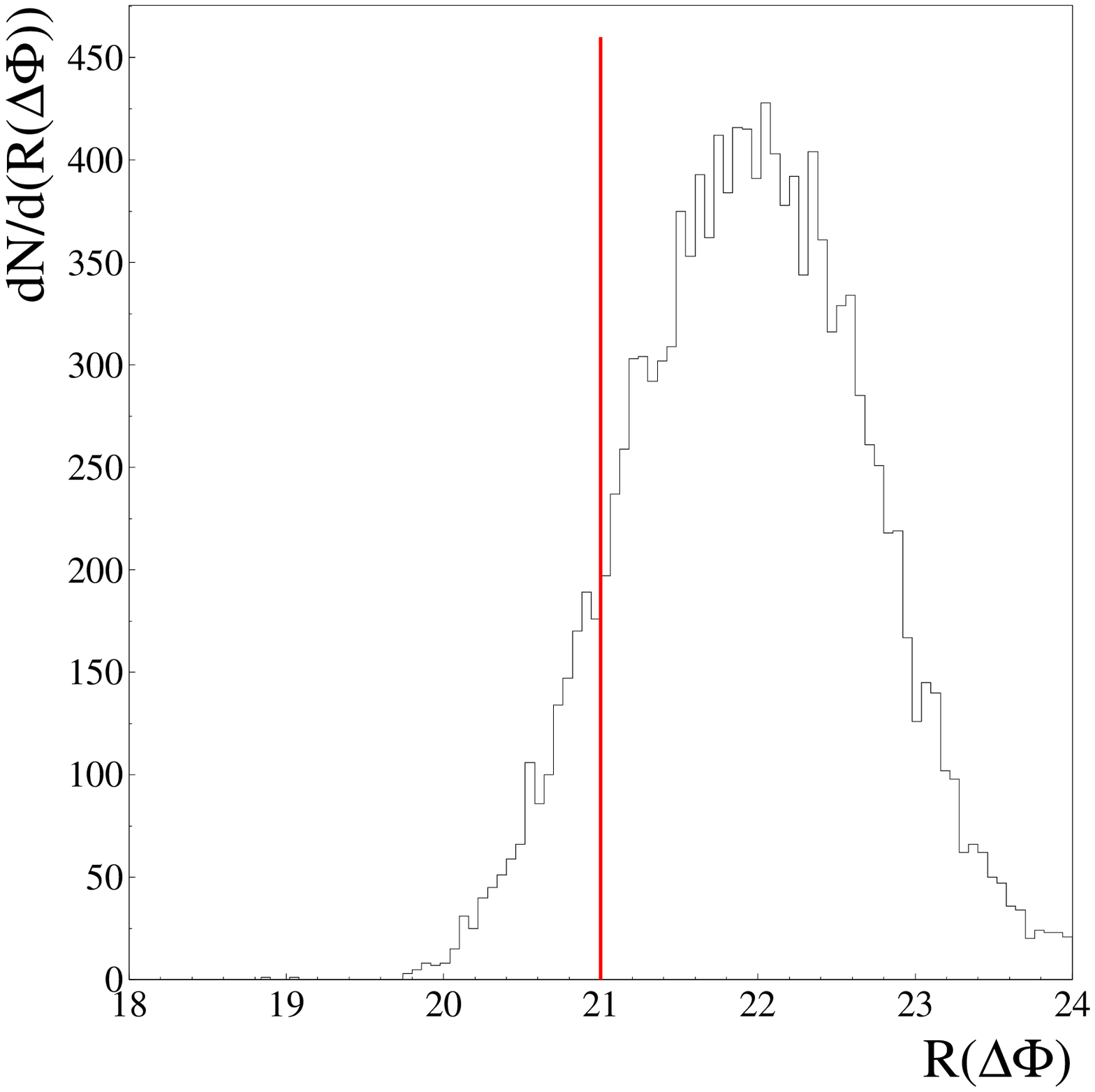}}
{\includegraphics[scale=0.33]{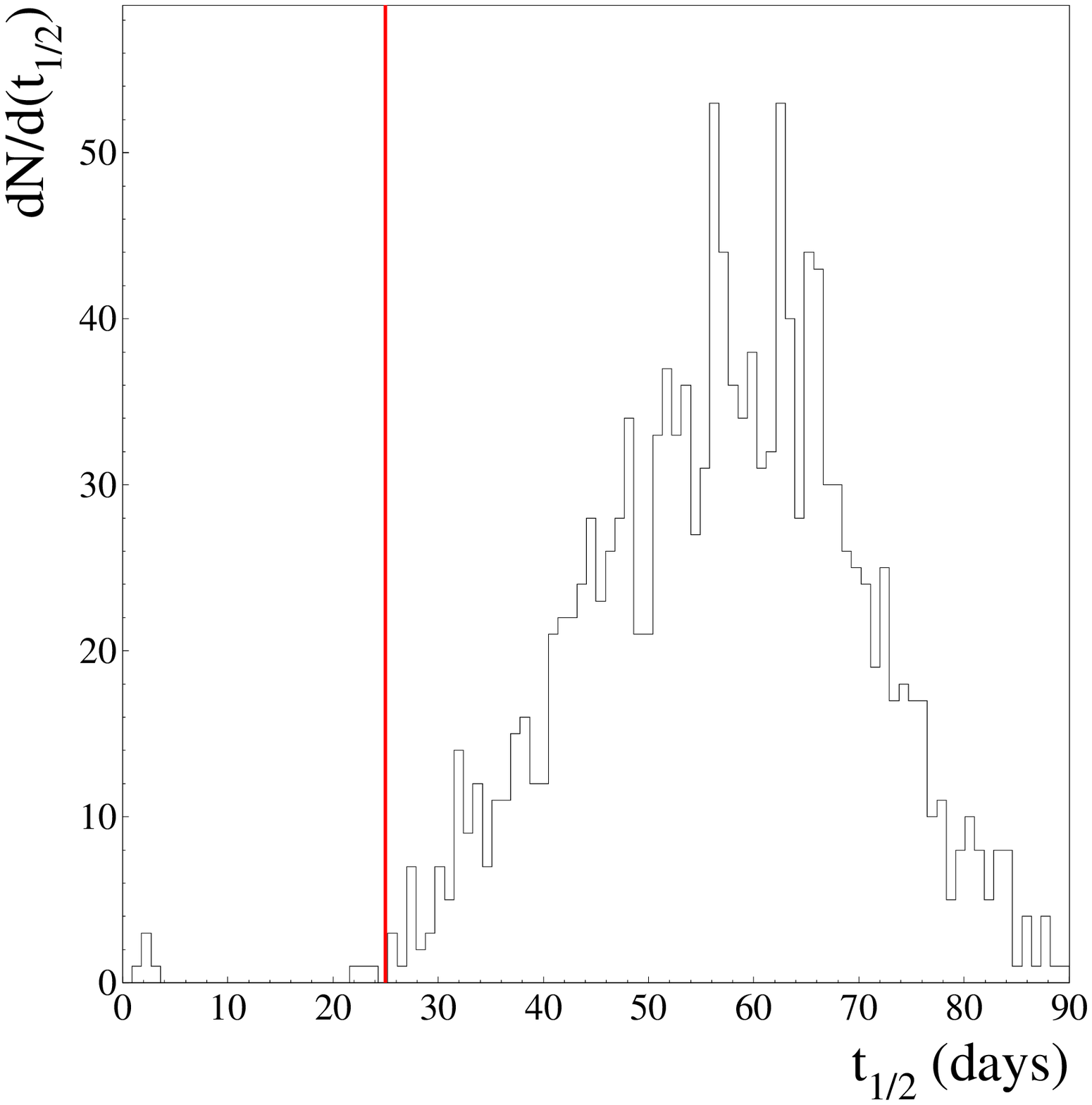}}
\end{center}

\caption{Top: Distribution of flux deviations at maximum
for the selected events after the sampling cut.
Bottom: Duration distribution for the selected events after the cut on $R(\Delta\Phi)$.}
\label{fig:tdemi}
 
\resizebox{\hsize}{!}{\includegraphics{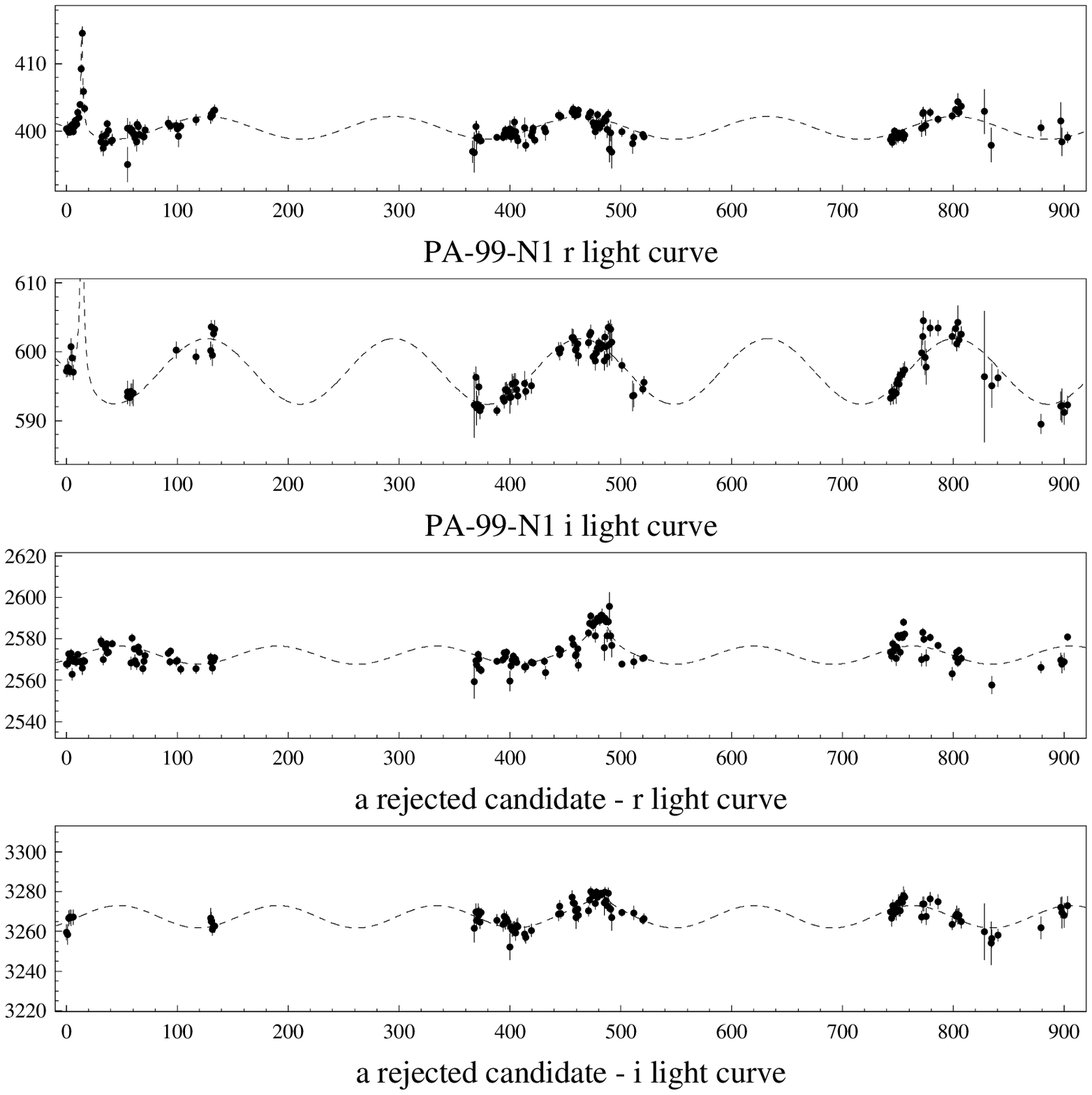}}
\caption{
$r$ and $i$ filter 3-year light curves for 2 selected
variations before the last cut. Upper panels, PA-99-N1, an accepted
candidate. Lower panels: a rejected candidate.
The dashed line is the best-fit for a \pacz\ bump with a sinusoidal background. 
The abscissae are time in days (JD-2451392.5). The ordinates  are flux in ADU/s.
}
\label{fig:paczsin}
\end{figure}

\begin{table*}[t]
\begin{center}
\begin{tabular}{|c|c|}
\hline
  criterion & number of selected light curves \\
\hline
\hline
  cluster detection ($Q>100$) & $\sim 10^5$ \\
\hline
\hline
  signal to noise ratio ($L_1>40$) and second bump ($L_2/L_1 < 0.5$) & $\sim 4 \cdot 10^4$ \\
\hline
  shape analysis  : $\chi^2/\textrm{dof} < 10$ (7 parameter \pacz\ fit) &  $\sim 3 \cdot 10^4$ \\
\hline
  time sampling along the bump & $\sim  10^4$ \\
\hline
\hline
  flux deviation: $R(\Delta\Phi)<21$  & $\sim 1.5 \cdot 10^3$ \\
  time width: $t_{1/2}<25$ days &  9 \\
\hline
  second bump analysis & 6\\
\hline
\end{tabular}
\caption{Summary of the selection criteria and number of the selected light curves.\label{tab:selection}}

\vspace{.3cm}

\begin{tabular}{|c||c|c|c|c|}
\hline
 & PA-99-N1 & PA-99-N2 & PA-00-S3 & PA-00-S4\\
\hline
\hline
$\alpha$ (J2000) & 00h42m51.19s & 00h44m20.92s & 00h42m30.27s &
00h42m29.98s\\
$\delta$ (J2000) & $41^\circ 23'56.3''$ & $41^\circ 28'44.8''$ & $41^\circ 13'00.6''$
& $40^\circ 53'46.1''$\\
$\Delta\Theta$ & $7'53''$ & $22'04''$ & $4'06''$ & $22'33''$\\
\hline
$t_{1/2}$ (days) & $1.83^{+0.12}_{-0.11}$ & $22.16^{+0.12}_{-0.12}$ & $2.303^{+0.074}_{-0.062}$ &
$1.96^{+0.09}_{-0.10}$\\
$R(\Delta\Phi)$ & $20.83\pm 0.10$ & $19.10\pm 0.10$ & $18.80\pm 0.20$ & $20.7\pm 0.20$\\
$V-R$ & $1.2\pm 0.2$ & $1.0\pm 0.1$& & \\
$R-I$ & & & $0.6\pm 0.1$ & $0.0\pm 0.1$\\
$t_0$ (JD-2451392.5) & $13.85\pm 0.05$ & $71.70\pm 0.10$ & $458.40\pm 0.02$ & $488.90\pm 0.07$\\
\hline
$t_{\rm E}$ (days) & $8.3^{+4.5}_{-2.7}$ & $71.1^{+4.1}_{-3.7}$ &
$10.4^{+2.5}_{-2.3}$ & $135^{+??}_{-76}$\\
$u_0$ & $0.070^{+0.046}_{-0.030}$ & $0.1014^{+0.0070}_{-0.0067}$ & $0.070^{+0.024}_{-0.017}$ &
$0.0042^{+0.056}_{-??}$\\
$\phi_{r}^*$ (ADU/s) & $1.17^{+0.76}_{-0.49}$ & $10.87^{+0.77}_{-0.83}$ & $8.9^{+3.3}_{-2.1}$ &
$0.11^{+0.15}_{-??}$\\
$\phi_{g}^*$ (ADU/s) & $0.35^{+0.24}_{-0.15}$ & $3.57^{+0.28}_{-0.25}$ & &\\
$\phi_{i}^*$ (ADU/s) & & & $11.7^{+4.0}_{-2.9}$ & $0.07^{+0.10}_{-??}$ \\
$A_\textrm{max}$ & $14.3^{+9.4}_{-6.1}$ & $9.9^{+0.68}_{-0.65}$ &
$14.3^{+4.9}_{-3.5}$ & $200^{+3200}_{-??}$\\
\hline
$\chi^2/$dof & 1.1 & 9.3 & 2.1 & 0.9\\
\hline
\end{tabular}
\caption{Main characteristics of the four already published microlensing
candidates. $\Delta\Theta$ is the projected separation from the centre of M31.
The magnitudes correspond to the maximum flux deviation and
are given in standard Johnson/Cousins  system. The results
reported here are the results of the \pacz\ fit alone, even when extra
information is available, as is the case for PA-99-N1 and PA-99-N2.
}
\label{tab:gold}
\end{center}
\end{table*}

Another crucial element in the selection is the choice
for the required sampling along the bump. In fact,
while a good sampling
is needed in order to meaningfully characterise the detected
variation, demanding too much in this respect could lead us
to exclude many \emph{bona fide} candidates.
Using the values of $t_{1/2}$ and $t_0$ determined in the preceding step, we define 4 time intervals around
the time of maximum magnification $t_0$: $[t_0-3\,t_{1/2},\,t_0-t_{1/2}/2],\,
[t_0-t_{1/2}/2,\,t_0],\,[t_0,\,t_0+t_{1/2}/2]$ and
$[t_0+t_{1/2}/2,\,t_0+3\,t_{1/2}]$.
As a fourth cut we demand that a minimum number of observing epochs
$n_{min}$ occur in each of at least  3 of these time intervals. Clearly
$n_{min}$ cannot be as large for short events as for long ones. We choose
$n_{min}=1,\,2$ and 3 for  $t_{1/2} < 5,\, t_{1/2} \in (5,15)$
and $t_{1/2} >15$ days, respectively.
Furthermore, neither of the external intervals should fall
at the beginning or end of one of the three seasons and at the same
time be empty.

The cuts described above reduce our sample of potential events to $\sim 10^4$, about one tenth
of the initial set of selected variations, but still mostly variable stars.

In this paper, we restrict attention to \emph{bright}
microlensing-like variations, in particular we demand
$R(\Delta\Phi)<21$, although the observed deviations extend
down to $R(\Delta\Phi)\sim 24$
(Fig. \ref{fig:tdemi}). This reduces our set
of candidates by another factor of $\sim$ 10.

The \MC\ (Sect. \ref{sec:MC}) predicts most of the microlensing
events to be rather short. On the other hand, the observed $t_{1/2}$ distribution
shows a clustering of long variations centred on $t_{1/2}\sim 60$ days,
most of which are likely to be  intrinsically variable objects,
and a much smaller set of short-duration variations
(Fig. \ref{fig:tdemi}).
We demand $t_{1/2}<25$ days, which leaves us with only
9 \pacz-like flux variations.

\begin{figure*}[tbh!]
\begin{center}
\includegraphics[width=16cm]{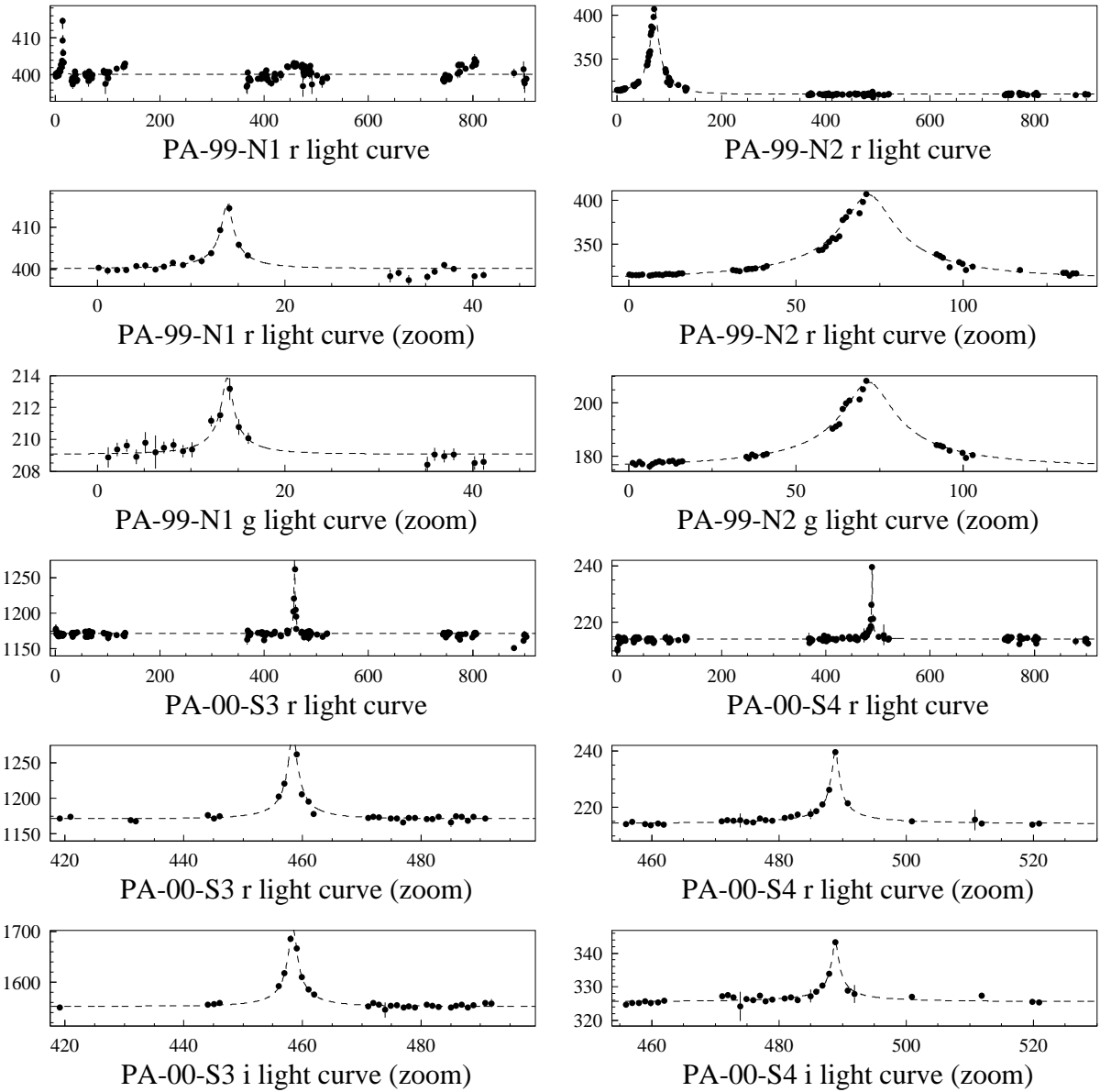}
\caption{
3-year light curves of the four microlensing events PA-99-N1, PA-99-N2,
PA-00-S3 and PA-00-S4. For each event, the top panel shows the whole light curve in the $r$
filter, while the 2 lower panels display zoomed light curves in all
bands for which data are available. Dashed lines are best-fit \pacz\ curves.
The abscissae are time in days (JD-2451392.5). The ordinates  are flux in ADU/s.
}
\label{fig:gold.cl}
\end{center}
\end{figure*}

Out of the 9 variations selected above, 5 show a significant second
bump. We want to exclude variable stars,
while keeping
real microlensing variations that happen
to be superimposed on a variable signal. For  most
variable stars  the secondary bump
should be rather similar but not identical to the detected one. To make
use of this fact we perform a three-colour
fit, modelling the light curve with a
\pacz\ bump plus a sinusoidal signal, and then
compare the time width and the flux variation of the sinusoidal part
with those of the \pacz\ bump.
Because our model is very crude and because we know that
variable stars may show an irregular time behaviour, we do not ask
for a strict repetition of the bump along the baseline
to reject a variation. We exclude a light curve if both
the $R(\Delta\Phi)$ difference between the two bumps
is smaller than 1 magnitude \emph{and} the time width
of the sinusoidal part is compatible
with that of the bump within a factor of 2.
Three out of nine variations are excluded
in this step. For all three the detected bump is relatively
long ($t_{1/2} >$ 20 days) and faint ($R(\Delta\Phi)>20.5$).
Furthermore, on the images the position of the second bump appears to be
consistent with that of the detected bump, clear
evidence in favour of the intrinsically variable origin of these variations.
Two other light curves are retained, although they show a significant secondary
bump; in both cases, the secondary bump is much longer than the main one.
Besides, in both cases the direct inspection on the images reveals that the position
of the second bump is different from that of the detected one.
In order to make clear the sense of the present criterion, we show (Fig. \ref{fig:paczsin})
the result of the \pacz\ fit superimposed over a sinusoidal background
for two variations. In the upper panels is an \emph{accepted}
candidate, for which the short and bright bump
at $t_0\sim 13$ (JD-2451392.5) is clearly distinct
from the underlying variable signal. In the lower panels is
a \emph{rejected} candidate. The \pacz\ signal
originally selected with peak at $t_0\sim 480$ (JD-2451392.5)
is clearly undistinguishable from the underlying variable background.

We are now left with our final selection of 6 light curves showing an
achromatic, short-duration and bright flux variation compatible with a
\pacz\ shape. We denote them PA-99-N1, PA-99-N2, PA-00-S3, PA-00-S4,
PA-00-N6 and PA-99-S7. The letter N(S) indicates whether the event lies in the
north (south) INT WFC field, the first number (99, 00, or 01) gives the
year during which the maximum occurs, and the second has been assigned
sequentially, according to when the event was identified.

In Table \ref{tab:selection} we report in sequence each step
of the pipeline with the number of the selected candidates
remaining.

\section{Microlensing events} \label{sec:events}

\subsection{POINT-AGAPE 3 years analysis results} \label{sec:pa-events}

\begin{figure}[tbh]
\begin{center}
\resizebox{\hsize}{!}{\includegraphics{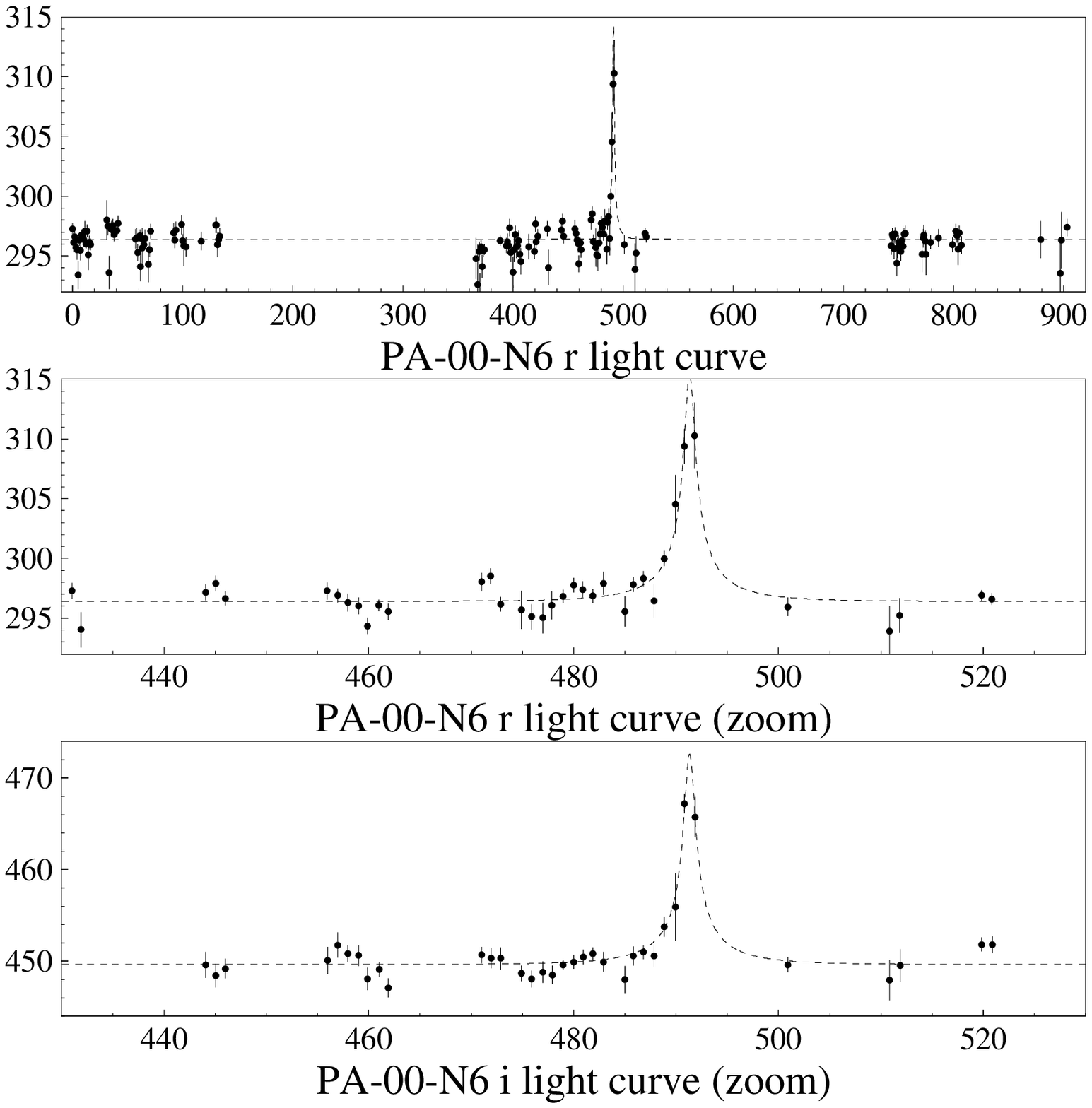}}
\caption{
3-year light curves of the microlensing event PA-00-N6.
Panels and symbols as in Figure \ref{fig:gold.cl}.
}
\label{fig:n6.cl}

\resizebox{\hsize}{!}{\includegraphics{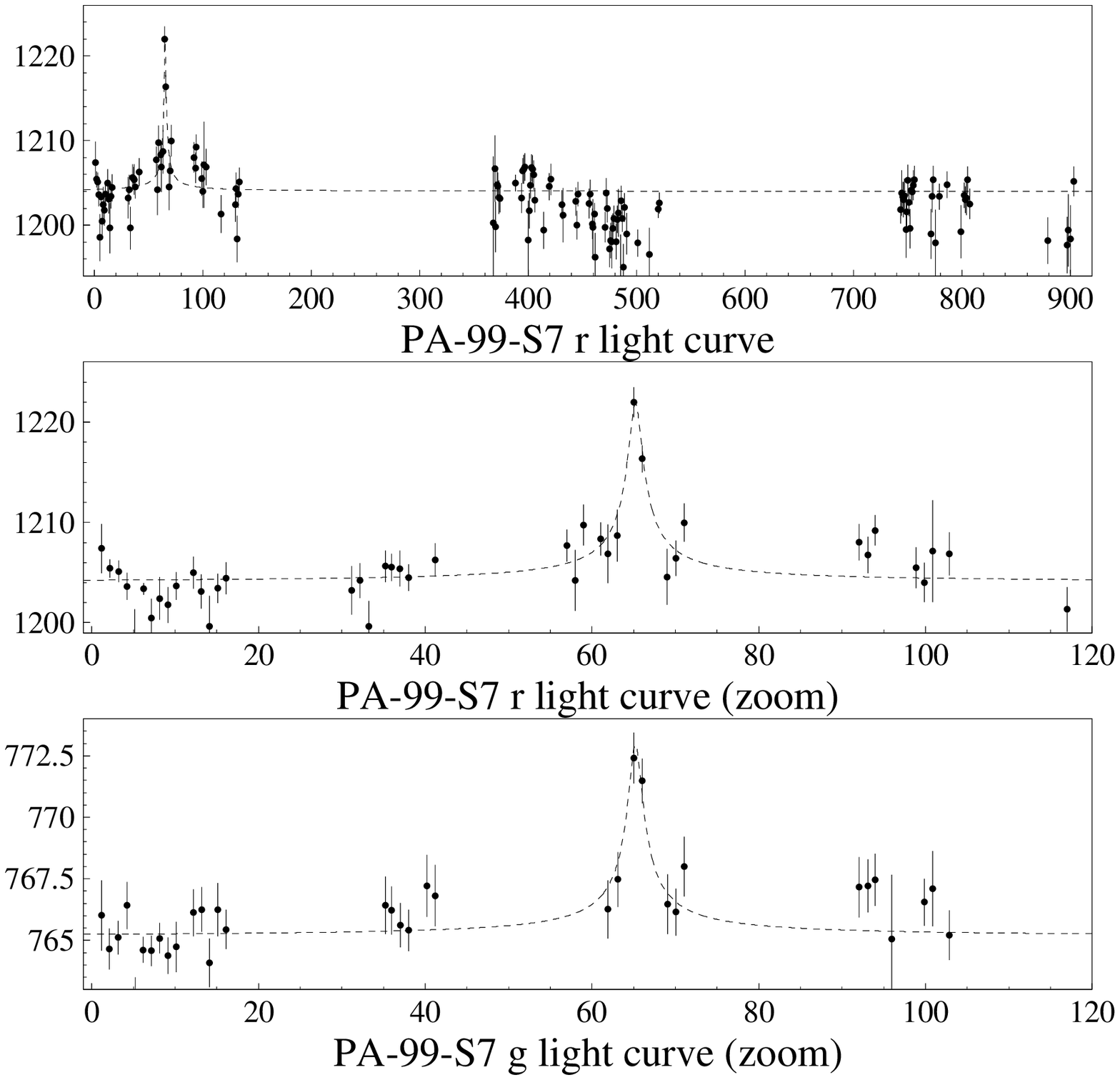}}
\caption{ 3-year light curves of the microlensing event PA-99-S7.
Panels and symbols as in Figure \ref{fig:gold.cl}. }
\label{fig:s7.cl}

\end{center}
\end{figure}

In this section we look at the 6 selected candidates in
detail. In Table \ref{tab:gold}  and Figure \ref{fig:gold.cl} we
recall the characteristics and light curves of the four already
published candidates\footnote{Full details can be found in
\citet{paulin02,paulin03,an04a}.}, while Table \ref{tab:new} and Figures
\ref{fig:n6.cl} and \ref{fig:s7.cl} are devoted to the two new
ones. The errors in $R(\Delta \Phi)$ and the colour index are
dominated by the uncertainty in the calibration of the observed
flux with respect to the standard magnitude system, except for
PA-00-N6. When the 7-parameter \pacz\ fit does not converge
properly, the time width and the flux increase are estimated from
a degenerate fit \citep{gould96}.

\begin{table}[tbh]
\begin{center}

\begin{tabular}{|c||c|c|}
\hline
 & PA-00-N6 & PA-99-S7\\
\hline
\hline
$\alpha$ (J2000) & 00h42m10.70s & 00h42m42.56s  \\
$\delta$ (J2000) & $41^\circ 19'45.4''$& $41^\circ 12'42.8''$ \\
$\Delta\Theta$ & $7'16''$& $3'28''$\\
\hline
$t_{1/2}$ (days) & $1.77^{+0.57}_{-0.60}$ & $4.10^{+0.85}_{-0.73}$\\
$R(\Delta\Phi)$ & $20.78^{+0.18}_{-0.31}$& $20.80 \pm 0.10$ \\
$V-R$ & & $0.79\pm 0.14$ \\
$R-I$ & $0.51^{+0.25}_{-0.43}$& \\
$t_0$ & $491.30 \pm 0.07$ & $65.21 \pm 0.14$ \\
\hline
$t_{\rm E}$ (days) & $8.3^{+10.5}_{-4.1}$ & -\\
$u_0$ & $0.07^{+0.13}_{-0.052}$ & -\\
$\phi_{r}^*$ (ADU/s) & $1.40^{+2.6}_{-0.95}$ & -\\
$\phi_{i}^*$ (ADU/s) & $1.7^{+3.2}_{-1.2}$ & -\\
$A_\textrm{max}$&$14^{+26}_{-11}$&-\\
\hline
$\chi^2/$dof & 1.0 & 1.3\\
\hline
\end{tabular}
\caption{Main characteristics of the two new microlensing
candidates. The parameters are the same as in Table \ref{tab:gold}.}
\label{tab:new}
\end{center}
\end{table}

\begin{figure}
\resizebox{\hsize}{!}{\includegraphics{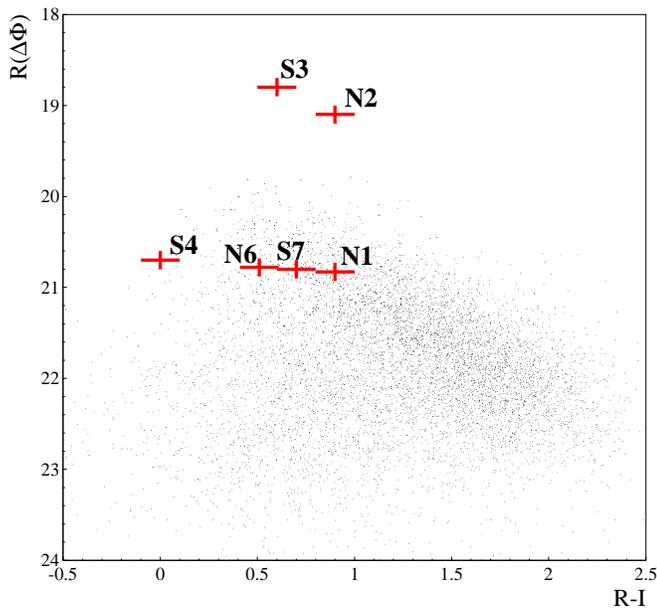}}
\caption{$R(\Delta\Phi)-(R-I)$ colour-magnitude diagram
for the $\sim 10\,000$ variations selected before the cut
on the flux deviation at maximum. Superimposed
we show the positions of the 6 selected candidates.
The $R-I$ colours for $PA-99-N2$ and $PA-99-S7$
are estimates derived from the observed $V-R$ colours.
}
\label{fig:colmag}
\end{figure}

The source star of PA-99-N1 has been identified on HST archival
images \citep{auriere01}. Fixing the source flux at the observed
values, $\phi^*_r=1.02\, \mathrm{ADU/s}$ and $\phi^*_g=0.28\,
\mathrm{ADU/s}$, we obtain $t_{\textrm E}=9.20\pm 0.61$ days and
$u_0=0.060\pm 0.005$, compatible within $1\sigma$  with the values
reported in Table \ref{tab:gold}, obtained from our data alone.
Finally, the HST data allow us to estimate the colour $(R-I)\sim 0.9$.
In \citet{an04a}, we have demonstrated that PA-99-N2, which shows
significant deviations from a simple \pacz\ form, is compatible
with microlensing by a binary lens. The binary-fit parameters are
characterised by a longer time scale and higher magnification than
the point-lens fit. In the best-fit solution we find $t_{\textrm
E}=125.0 \pm 7.2$ days, $u_0=(3.60 \pm 0.37)\times 10^{-2}$,
$\phi^*_r=4.76\pm 0.34$ ADU/s, and a lens mass ratio $\sim 1.2
\times 10^{-2}$. Under the assumption that the lens is associated
with M31 (rather than the MW), the lower bounds on the
angular Einstein radius ($\theta_{\rm E}>25\,\mu\rm as$)
deduced from the absence of
detectable finite-source effects implies that the source-lens
relative velocity is $v_\perp > 280$ km/s, and the source-lens
distance is $d_{ls}>45\,{\mathrm{kpc}}(M/M_\odot)^{-1}$, where $M$ is
the lens mass.  These facts, together with PA-99-N2's large
distance from the M31 centre ($\sim 22'$) make it very unlikely to
be due to an M31 star, while the prior probability that it is due
to a MW star is extremely low. Hence, PA-99-N2 is a very strong
MACHO candidate (either in M31 or the MW). The sampling and the
data quality along the bump are also good enough  to permit a
reliable estimate of all 7 parameters of the \pacz\ fit for the
event PA-00-S3. For PA-00-S4 we obtain only a reliable lower limit
on $t_{\textrm E}$, and accordingly an upper limit on $u_0$, as
indicated by the question marks in Table \ref{tab:gold}.

For PA-00-N6, the data allow us to evaluate the
full set of \pacz\ parameters. Note the rather short Einstein time,
$\sim 10$ days, similar to those of PA-99-N1 and PA-00-S3.

As in the case of PA-99-N1 \citep{paulin03},
PA-99-S7 lies near (within 4 pixels) of a long-period red
variable star. This induces a secondary bump, which is particularly visible in
the $i$ light curve. PA-99-S7 has been accepted by the last step of
our selection pipeline, despite this second bump being responsible
for poor stability of the baseline. In this case, the data do not allow us
to break the degeneracy among the \pacz\ parameters and therefore do not allow a reliable
estimate of the Einstein time.

A colour-magnitude diagram of the $\sim 10\,000$ variations selected
after the sampling cut is shown in Figure \ref{fig:colmag}. Superimposed
we indicate the position of the 6 variations finally selected after all cuts.
In particular, we note the peculiar position
of PA-99-N2, which (together with PA-00-S3) is unusually
bright relative to the other variations.
Recall that PA-99-N2 is also the longest selected variation,
with $t_{1/2}\sim 22$ days.
As we have already excluded short-period variables,
the sample shown is dominated
by red, long-period variables of the Mira type
with $R(\Delta\Phi)>21,\,(R-I)>1$.
For a detailed discussion of the variable star populations detected within
our dataset see \cite{an04b}.

The spatial position for the detected events projected
on the sky  is shown, together with the INT fields, in Figure \ref{fig:field}.
Note the two new events are located within a rather small
projected distance of M31's centre.

\subsection{Variable Contamination \label{sec:varcon}}

Probably the biggest single problem in the interpretation of
microlensing events drawn from faint sources is the possibility
that the sample may be contaminated with rare variables.
For relatively bright sources, such as those being detected
by the thousand toward the Galactic bulge \citep{udalski03},
microlensing events are easily distinguished from variables
by their distinct shape.  However, as the S/N declines,
such discrimination becomes more difficult.  Experiments
toward the LMC provide sobering confirmation of the
legitimacy of this concern.  Both of the original microlensing
candidates reported by the EROS collaboration \citep{aubourg93}
were subsequently found to be variable stars, while some candidates
found by the MACHO collaboration \citep{alcock97,macho00} were
also subsequently recognized as possible or certain variables.
The SuperMACHO collaboration \citep{becker04}, which probes
about 2 mags fainter than MACHO or EROS in its microlensing search toward
the LMC, has so far found it extremely difficult to distinguish between
genuine microlensing events and background supernovae (C.~Stubbs 2005,
private communication).  Thus, when reporting a handful of microlensing
candidates drawn from 3 years of monitoring of a large fraction of an
entire L* galaxy, we should cautiously assess the possibility of
variable contamination.

If variables were contaminating our sample, they would have to reside
either in the MW or in M31 itself, or they could be background supernovae.
We consider these locations in turn.

There are three arguments against MW variables: distribution on the sky,
absence of such variables in the Galactic microlensing studies, and
lack of known classes of Galactic variables that could mimic microlensing.
First, of the 5 microlensing candidates that enter our event-rate analysis
(i.e., excluding the intergalactic microlensing candidate PA-00-S4),
4 lie projected in or near the M31 bulge.  This strongly argues that
they are, in their majority, due to M31 sources, which are also
heavily concentrated in this region.  By contrast, Galactic variables
would be spread uniformly over the entire field.  Of course, this
does not rule out the possibility of minor contamination by such variables.

However, if there were a class of variables that could even weakly
mimic short microlensing events with flux variations corresponding to
$R(\Delta\Phi)<21$, then these would have easily shown up in Galactic microlensing
experiments.  For example, the OGLE-III microlensing survey covers over
50 deg$^2$ toward the Galactic bulge, more than 100 times larger
than our survey toward M31.
The OGLE survey does not go as deep as ours because their telescope is
smaller (1.3m) and their exposure times are shorter (2 min), although
these factors are somewhat compensated by their denser temporal coverage.
Ignoring this shallower depth for the moment, and restricting
consideration to $\la 3\,\rm kpc$ (where most of our foreground
MW disc stars lie) the projected density of disc stars is about 10 times higher in the OGLE
fields than in ours because they lie at lower Galactic latitude.  Hence,
one would expect of order 1000 times more such variables to appear
in the OGLE fields than in ours.  Of course, the majority of these
would be  $R(\Delta\Phi)\sim21$ and so of such low signal-to-noise ratio that they would
not appear as OGLE candidates, or if they did, would
escape recognition as variables.
However, $\sim 1/125$ would lie 5 times closer
and so be 3.5 mag brighter, i.e., $R(\Delta\Phi)<17.5$, corresponding to $I\la 17$,
and these would have good signal-to-noise ratio.  No such variable
population is reported.  A similar argument applies to Galactic halo
stars, which would also be much denser in the OGLE-III fields than in ours.

Third, there are no known candidate classes of Galactic variables that
could mimic the M31 microlensing events.  The one possibility is
dwarf-novae, which have been reported as faking microlensing events
toward the LMC \citep{eros2} and M22 \citep{bond05}.  However, with
typical peak absolute magnitudes of $M_V\sim 2$ \citep{warner95}, they
would have to lie well outside the Galaxy to appear as $R(\Delta\Phi)\sim21$
fluctuations.

While the case against M31 variables is not as airtight as against Galactic
ones, it is still quite strong.  The basic argument is that if the
sources are in M31, then they must suffer luminosity changes corresponding
to $M_R<-3.5$ on quite short timescales ($t_{1/2}<5\,$days for all
candidates except PA-99-N2).  There are no known classes of variables
that do this except for novae.  However, novae show brighter variations
and strongly asymmetric light curves characterized by slow descents
(a selection of novae variations in our dataset is discussed in \citealt{darnley04}).
While in principle our microlensing candidates could be due
to some new, so far unrecognized (nor even conjectured) type of stellar
variability, the great brightness and very short timescale of the
observed events impose severe restrictions on candidate mechanisms
of variability.

Novel mechanisms to explain the sixth event, PA-99-N2, would be less
constrained because it is much longer, $t_{1/2}\sim 22\,$days.
However, being long as well as very bright ($R(\Delta\Phi)\sim 19$),
its signal-to-noise ratio is quite high.  This permits us to check
for achromaticity with very good precision.  Even the deviations
from a simple Paczy\'nski shape are achromatic and can be reproduced
by a binary-lensing curve \citep{an04a}. That is, PA-99-N2 is an
excellent microlensing candidate on internal evidence alone.

Finally, we remark on supernovae which, as noted above, plague the
SuperMACHO project and also were a difficult contaminant for the
MACHO and EROS projects.  There are two principal arguments against
supernovae.  First, the FWHMs of all but one of the events are too short
for supernovae while, as we have just argued, the sixth event is
achromatic and fit by a binary-lens light curve and therefore almost
certainly microlensing.  Second supernovae cannot be responsible for the
majority of the events because the supernovae would be uniformly
distributed on the sky while the actual events are highly clustered
near the centre of M31.

For completeness, we address one other concern related to variability:
the possibility that the source displays a signature
of variability away from the microlensing event. In this case,
one might worry that this ``event'' is actually an outburst
from an otherwise low-level variable.  Recall that our selection
procedure actually allows for a superpixel to show lower-level
variability in addition to the primary ``event'' that is characterized as
microlensing, and to still be selected as a candidate.  This is necessary
because about 15\% of pixel light curves within $8'$ of the M31 centre
(a region containing most of our events) show variable-induced ``bumps''
with likelihood $L_1>40$.  So we would lose 15\% of our sensitivity
if we did not try to recover microlensing events with such secondary
bumps.  One event (PA-99-N1) out of four in this region displays such
a severe secondary bump.  This 25\% rate is within Poisson uncertainties
of the 15\% expectation.  In addition, a second event (PA-99-S7) displays a
secondary bump at less than half this threshold.

It must be stressed, however, that through a Lomb analysis we find
that neither of the \emph{source stars} for these two events shows any sign
of variability apart from the microlensing event.
In both cases, the source of the lower-level variation lies several
pixels from the microlensing event.

In brief, while we cannot absolutely rule out non-microlensing sources
of stellar variability, all scenarios that would invoke variability
to explain our candidate list are extremely constrained, indeed contrived.

\subsection{A likely binary event} \label{sec:pa-s5}
Our selection pipeline is deliberately biased to reject flux variations
that strongly differ from a standard \pacz\ light curve. In
particular, it cannot detect binary lens events with caustic
crossing. We discuss here a blue flux variation ($R-I\sim 0$) that failed to pass the
$\chi^2$ cut, but is most probably a binary lens event: PA-00-S5. The light
curve, which involves a
short ($t_{1/2} \sim 2 \mbox{ days}$) and bright peak  followed by
a plateau,  is suggestive of binary lensing with a caustic crossing.
The photometric follow-up of this event is tricky,
particularly in the $i$ band, because a faint resolved red object
lies about 1.5 pixels away. To overcome this difficulty, we have
used  a more refined difference image photometry that includes modelling
the PSF.

\begin{figure}[ht]
\resizebox{\hsize}{!}{\includegraphics{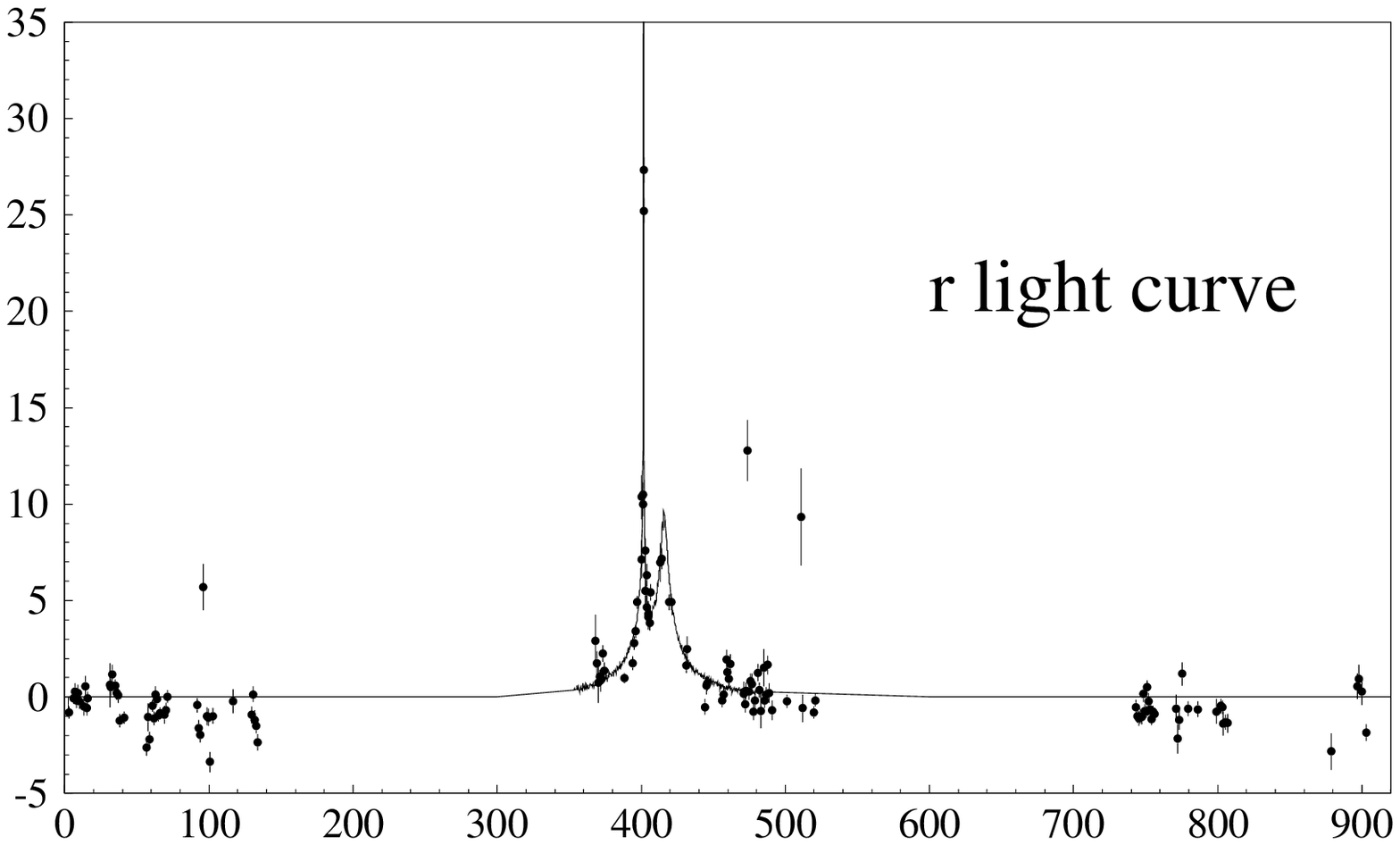}
\includegraphics{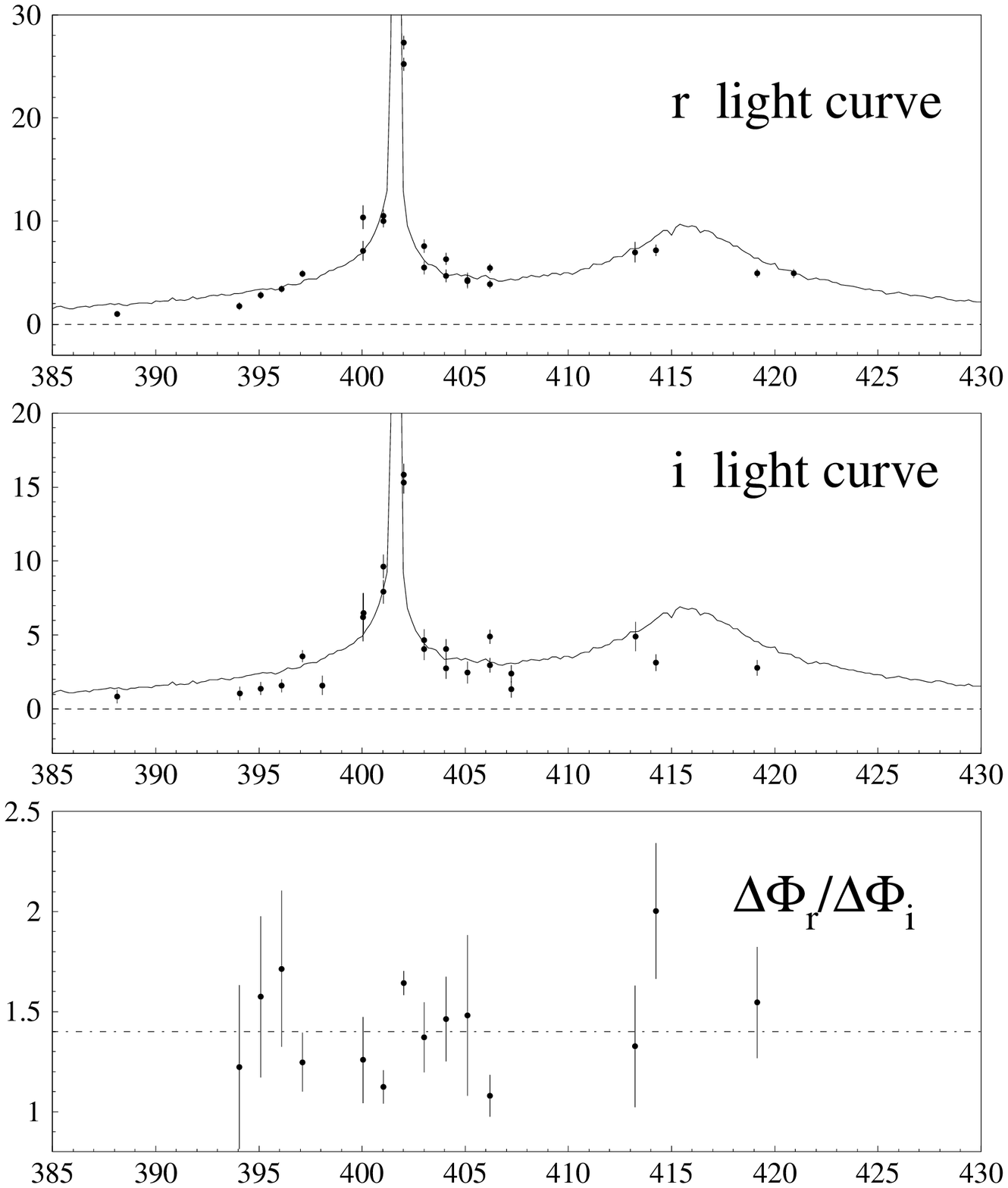}}
\caption{A binary solution superposed on the differential-photometry
light curve of the binary-lens candidate PA-00-S5.
Upper panel: full $r$ light curve; middle panels: $r$ and $i$ zooms
around the bump region, the dotted line shows the baseline; bottom
panel: the colour ratio $\Delta\Phi_r/\Delta\Phi_i$, the dash-dotted
line being the average colour ratio. The abscissae are time in days
(JD-2451392.5), the ordinates of the three upper panels are flux in ADU/s.
\label{fig:s5}
}
\end{figure}

We have found a binary lensing solution
that convincingly reproduces the shape of
the bump.
  The corresponding light curve, superimposed on
  the data obtained using difference image photometry, is displayed in Figure
  \ref{fig:s5}, where we show the full $r$ light curve, zooms of the
  bump region in the $r$ and $i$ bands, and the ratio of flux increases
  $\Delta\Phi_r/\Delta\Phi_i$.

 This solution is a guess, neither optimised nor checked for uniqueness.
The parameters are as follows: the distance between the two masses
is $d=0.63$ in unit of the Einstein radius $R_E$, the mass ratio
is $q=1/2$; the distance of closest approach to the barycentre,
$u_0=0.17$, is reached at $t_0=411 \mbox{(JD-2451392.5)}$; the
Einstein time scale is $t_E=50 \mbox{ days}$; the source crosses
the binary axis at an angle of $58.5^\circ$, outside the two
lenses and close to the heavy one.

The location of PA-00-S5 is $\alpha$ = 00h41m14.54s, $\delta=40^\circ
48'37.7''$, J2000, some $32'$ away from M31's centre.
This event cannot enter the discussion of the following sections
because it does not survive our full selection pipeline and because
the possibility of caustic crossings is not included in the
simulation. Nevertheless, if this event is due to microlensing, the
lens is most probably a binary MACHO.

\subsection{Comparison with other surveys} \label{sec:comp}
The first microlensing candidate reported in the direction of M31,
AGAPE-Z1, was detected in 1995 by the AGAPE collaboration
\citep{agape99}. AGAPE-Z1 is a very bright event, $R = 17.9$, of
short duration, $t_{1/2}$ =5.3 days, and located  in the very central
region of M31, at only $\sim 42''$ from the centre.

The MEGA collaboration has  presented
results from a search for microlensing events
using the first 2 years of the same 3-year data set
analyzed here \citep{dejong04},
but  a different technique.
In contrast to the present analysis, they do not impose
any restriction on $t_{1/2}$ and $R(\Delta\Phi)$.
As a result, they select 14 microlensing candidates.
All of them belong to our initial catalogue of flux variations. However,
beside MEGA-7 and MEGA-11 (corresponding to PA-99-N2 and PA-00-S4, respectively),
the remaining 12 flux variations are fainter than allowed by our magnitude cut
($R(\Delta \Phi) < 21$). Moreover, MEGA-4, MEGA-10, MEGA-12 and MEGA-13
have time widths longer than our threshold of 25 days.

The WeCAPP collaboration, using an original set of data acquired
in the same period as our campaign, reported the detection of two
microlensing candidates \citep{riffeser03}. The candidate
WeCAPP-GL1 is PA-00-S3. We did not detect the candidate WeCAPP-GL2
(short enough but probably too faint to be included in our
selection) because its peak falls in a gap in our observations.

The NAINITAL survey has recently reported \citep{nainital04} the
discovery of a microlensing candidate toward M31, quite bright
($R(\Delta\Phi)=20.1$) but too long ($t_{1/2}\sim 60$ days)
to be selected within our pipeline.

Recently we have reported \citep{belokurov05}
the results of a search for microlensing events obtained using a different
approach. Starting from
a different catalogue of flux variations
and using a different set of selection criteria
(in particular, we did not include any explicit cut
in $t_{1/2}$ or $R(\Delta\Phi)$),
we reported 3 microlensing candidates:
PA-00-S3, PA-00-S4 and a third one, which is not included
in the present selection. It is a short, bright, rather blue flux variation
($t_{1/2} =4.1$ days, $R(\Delta\Phi) = 19.7$,
$R-I = 0.0$), detected in the third year  ($t_0=771$ (JD-2451392.5)).
In the present analysis it is rejected because it fails to pass the
sampling cut: it does not have enough points on the rising
side to safely constrain its shape.
The position of this event, ($\alpha$=00h42m02.35s,
$\delta = 40^\circ 54'34.9''$, J2000), rather far away from the centre of M31
($\Delta\Theta = 22'59''$), is consistent with its being a MACHO
candidate. However, because it does not survive the present selection
pipeline, we do not include it in the following discussion.
A further analysis in which we follow a still different approach
is currently underway \citep{london}.

\section{The Monte Carlo analysis} \label{sec:MC}

The \MC\ attempts, for a given astrophysical context, to predict the
number of events expected in
our experiment, trying to mimic  the  observational
conditions and the selection process. Because these
can only partially be included in the \MC, the full simulation of our
observation campaign must involve the detection efficiency analysis
which is described in Sect. \ref{sec:eff}.

\subsection{The astrophysical model}

\subsubsection{The source stars}
Source stars are drawn according to the target M31
luminosity profile as modelled by \citet{kent89}. The 3-dimensional
distribution of bulge stars is also taken from \citet{kent89}.
The distance $z$ of disc stars to the disc plane follow a
$1/\cosh^2(z/H)$ distribution with $H=0.3\,{\mathrm{kpc}}$ as proposed by \citet{kerins01}.

\begin{figure}[h]
\includegraphics[scale=0.40,clip=true]{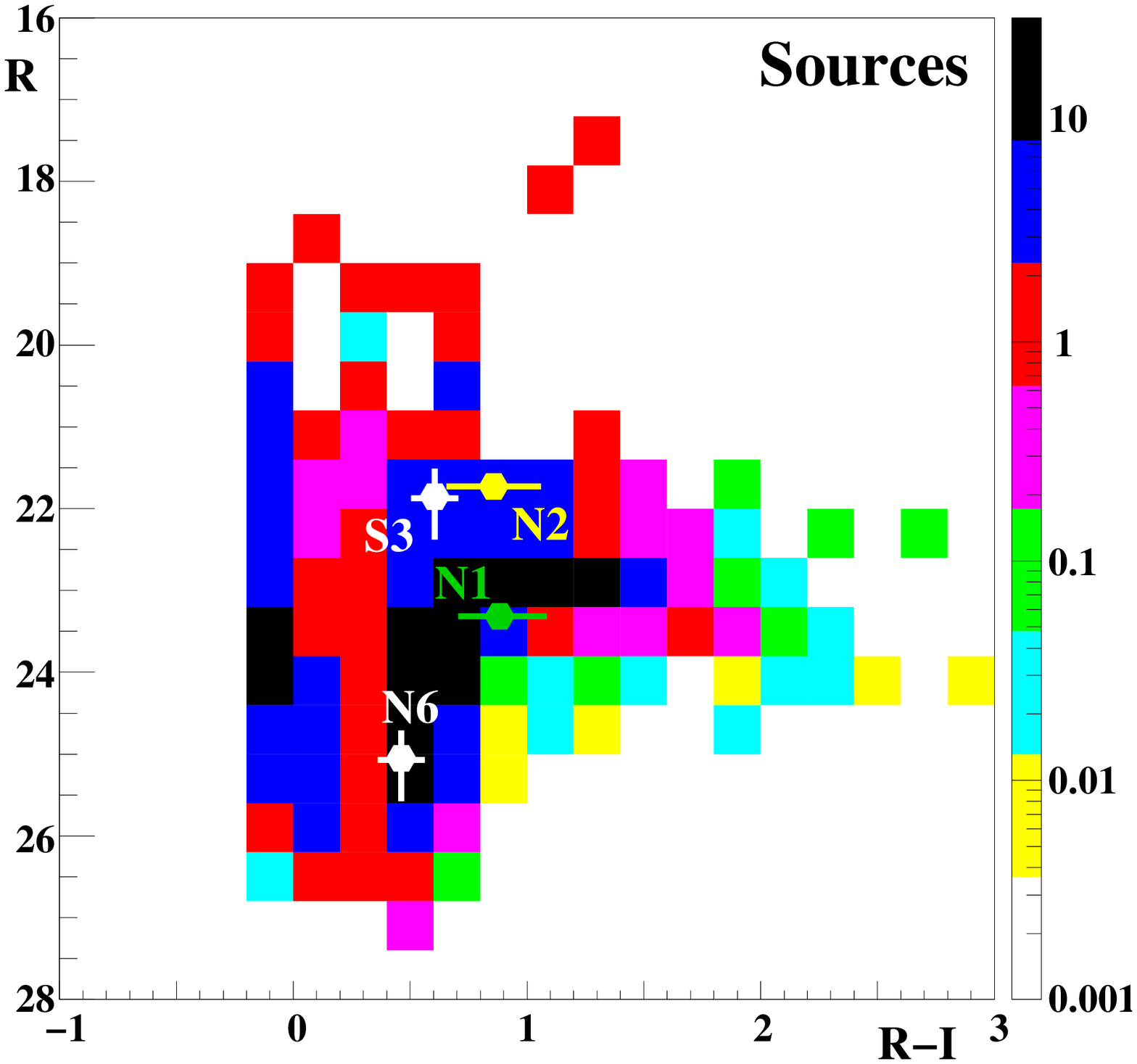}
\includegraphics[scale=0.40,clip=true]{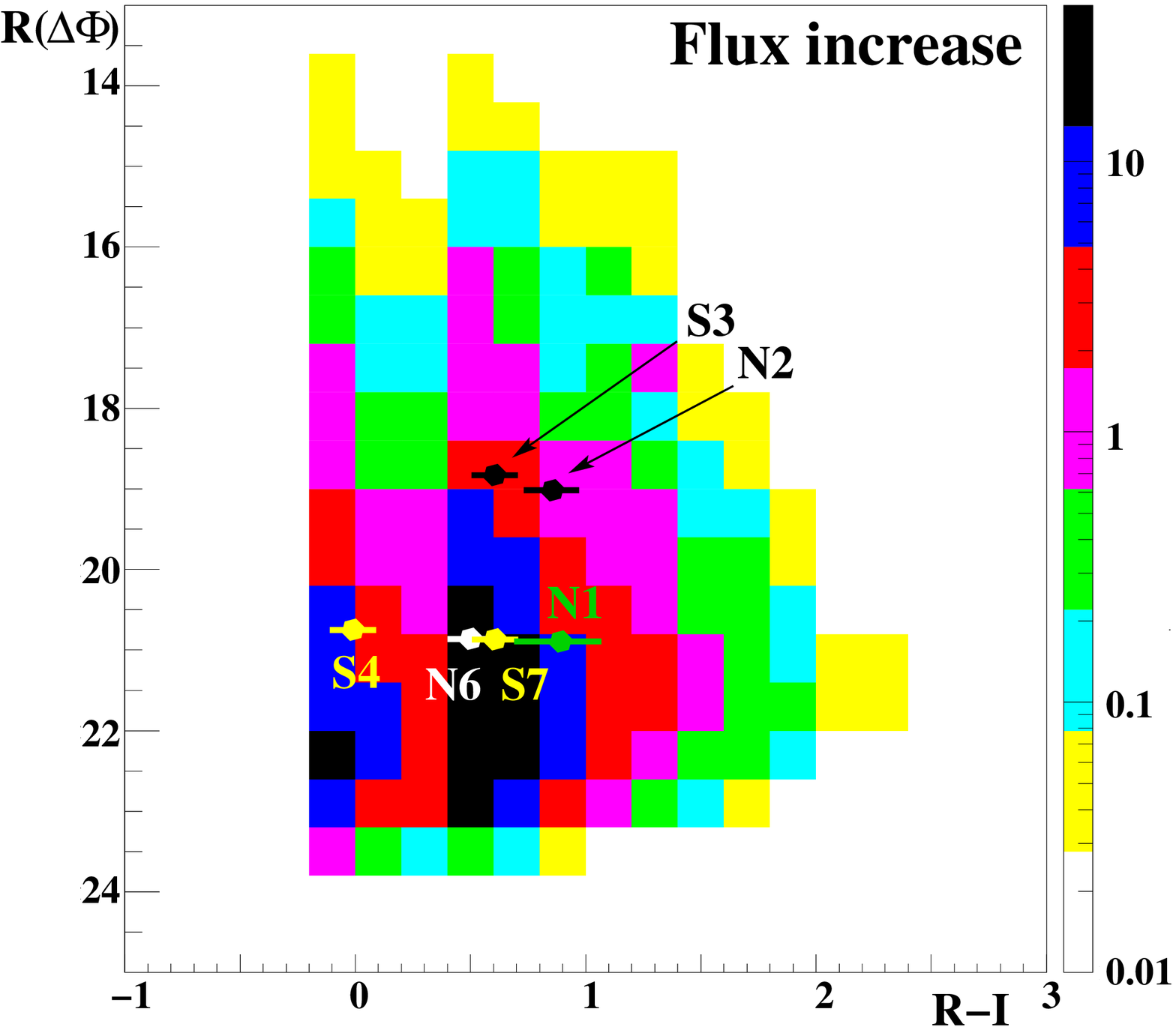}
\caption{The colour/magnitude event density
  distribution predicted by the \MC. Top panel: R
  magnitude of the source star. Bottom panel: R magnitude
  of the flux increase. The observed events are superposed on the
  diagrams. Only those events for which the source magnitude can be
  reliably extracted appear in the left panel. The colour scale
  shows the event density (in arbitrary units).
\label{fig:CM}
}
\end{figure}

The colour-magnitude distributions of disc and bulge stars are supposed
to have the characteristics of the Milky Way disc and bulge populations.
The distribution of disc stars is taken from the solar neighbourhood
data obtained by Hipparcos \citep{perryman97},
corrected at the bright end for the completeness
volume\footnote{The luminosity   function obtained in this way fully
  agrees with that presented in \citet{hipp97}.}   and incorporating at
low luminosity (needed for normalisation)
a Besan{\c c}on disc model \citep{robin03}. For the bulge we again use a
Besan{\c c}on model \citep{robin03} completed at the faint
end using \cite{han96}.
We construct two distinct types of ``colour-magnitude diagrams''
(CMDs) from the \MC\ and show these in Figure \ref{fig:CM} with
the position of the actual detected events superposed.
The first is a standard CMD, which plots apparent magnitude
versus colour for the sources of all the simulated microlensing
events that meet our selection criteria. In fact, however,
while the colours and magnitudes of all selected-event sources
are ``known'' in the \MC, they cannot always be reliably extracted
from the actual light curves: the colours are well-determined,
but the source magnitudes can only be derived from
a well-constrained \pacz\ fit (while some events have only
degenerate fits). We therefore also show in Figure \ref{fig:CM}
a second type of CMD, in which the ordinate is the magnitude
corresponding to maximum flux increase during the event ($R(\Delta \Phi)$).
It is always well-determined in both the \MC\ and the data.

To take into account the effect of the finite size of stars, which can
be important for low mass MACHOs, we have to evaluate the source
radii. To this end, we use a colour temperature relation evaluated
from the models of \citet{robin03}, and we evaluate the radii from
Stefan's law using a table of bolometric corrections from \citet{murdin01}.

We did not take into account possible variations of the interstellar
extinction across the field, although there are indications of
higher extinction on the near side \citep{an04b}. The best
indicator we have of  differential extinction is the asymmetry of
the surface brightness map, and this gives a flux attenuation by dust
on the near side of about 10\%. This is also the order of magnitude
of the average extinction one would obtain assuming that the M31 disc
absorption is about twice that of the MW disc. Indeed, as dust is confined
in a thin layer, extinction only significantly affects the stars on the
back side. Clearly an attenuation of about 10\% would not
significantly affect the results presented here.

\subsubsection{The lenses} \label{sec:lenses}
The lenses can be stars or halo objects, with the latter
being referred to as ``MACHOs''.
The stellar lenses can be either M31 bulge
or disc stars\footnote{We do not include lensing of M31 objects by stars
of the MW disc. This can be at most of the same order of magnitude
as M31 disc-disc lensing, which is included but turns out to be
small.}.

In the case of the bulge, we shall consider the microlensing contribution
of bulge stars with a standard stellar mass-to-light ratio.
Such models form the only true litmus test for whether or not dark matter
must be invoked, since the dark matter solution is classically required
to explain observations which cannot be accounted for by known populations.
The only dynamical requirement for our stellar bulge models is that their
dynamical contribution does not exceed the observed inner rotation curve.
They \emph{do not} need to fully reproduce the inner rotation curve,
though their failure to do so must be seen as evidence in itself for dark matter.
We shall from here onwards use the term \emph{stellar bulge}
to denote the contribution to the bulge from ordinary stars.
We use the term \emph{bulge} by itself to mean the entire dynamical bulge mass,
which must include the stellar bulge but which may also comprise additional mass
from unknown populations. We implicitly assume that the total bulge mass is fixed
by the rotation curve. We set out here to discover whether or not the rate
predicted by known stellar bulge and disc populations can feasibly account
for our observed microlensing candidates.

\paragraph*{Disk stellar lenses}~~    The disc mass
distribution is the same as in \citet{kerins01}:
$$
\rho =\rho_0 \exp\left(-\frac{r}{h}\right)/\cosh^2\left(\frac{z}{H}\right)
$$
with $\rho_0=0.3\,M_{\odot}\,{\mathrm{pc}}^{-3}$, $H=0.3\,{\mathrm{kpc}}$ and $h=6.4\,{\mathrm{kpc}}$.

The mass of the disc is $3\,10^{10}\,M_\odot$, corresponding to an
average  disc  mass-to-light ratio ${M/L}_B$ about 4.

\paragraph*{Bulge stellar lenses}~~ The bulge 3-dimensional mass distribution is
taken to be proportional to the 3-dimensional luminosity distribution, which
means that the bulge  ($M/L$) ratio is position independent.
Assuming that the M31 stellar bulge is similar to that of the Milky Way,
one can estimate from \citet{han-gould03} that $M/L_B \sim 3$ and
that it cannot exceed 4 (corresponding to bulge masses of 1.5 and $2\,10^{10}$ M$_\odot$
within 4 kpc). This can also be inferred by
combining results from \citet{zoccali00} and
\citet{roger86}. \citet{han-gould03} have shown that this
\emph{stellar} $M/L$ accurately predicts the optical depth that is
observed toward the MW bulge.

Estimates higher than the above values for the total bulge and disc $M/L_B$
have been quoted on dynamical grounds
\citep{kent89,kerins01,{baltz03},widrow03,geehan05,widrow05} and used
to make predictions on self lensing (e.g. \citealt{baltz03}).  In these
dynamical studies a heavy bulge ($M \sim 4\,10^{10}\,M_\odot$,
$M/L_B\sim 8$) is typically associated with a light disc
($M \sim 3\,10^{10}\,M_\odot$, $M/L_B \sim 4$), whereas a light bulge
($M \sim 1.5\,10^{10}\,M_\odot$, $M/L_B \sim 3$) goes with a
heavy disc ($M \sim 7\,10^{10}\,M_\odot$, $M/L_B \sim 9$). As stated above, such
large $M/L_B$ ratios mean that some kind of dark matter must be present
as no known ordinary stellar populations can provide such high $M/L_B$ ratios.
We shall refer to
these solutions to evaluate upper bounds on the self-lensing
contribution in Sect. \ref{sec:results}.

The stellar mass function is taken from \citet{kerins01} :
\beq \label{eq:imf}
\frac{dN}{dm} \propto \left\{\begin{array}{ll}
m^{-0.75} & (0.08\, M_\odot < m < 0.\,5 M_\odot)\\
m^{-2.2}  & (0.5\, M_\odot < m < 10\, M_\odot)
\end{array}\right.
\eeq
The corresponding average stellar mass  is $<m> \sim 0.65$ M$_\odot$.
We have also considered steeper mass functions,
as proposed by \citet{zoccali00}, for which
$<m> \sim 0.55$ M$_\odot$,
or by \citet{han-gould03}, for which $<m> \sim 0.41$ M$_\odot$. Our
results turn out to be rather insensitive to this choice.

\paragraph*{Halo lenses (MACHOs)}~~The MW and M31 halos are modelled
as spherical nearly isothermal distributions with a
core of radius $a$ :
\beq
\rho(r)=\frac{\rho_0\, a^2}{a^2+r^2} \label{rhohalo}
\eeq
The central halo density is fixed, given the core radius, to produce the asymptotic disc
rotation velocity far from the galactic centre. For the Milky Way the core radius
$a_{MW}$ is chosen to be $5\, \mathrm{kpc}$.
For M31 we choose $a_{M31} = 3 \,{\mathrm{kpc}}$ for our reference model but we have also
tried $a_{M31} = 5 \,{\mathrm{kpc}}$. A larger value for the core radius decreases the
number of expected events and makes their spatial distribution
slightly less centrally concentrated.

As nothing is known about the mass function of putative MACHOs, we
try a set of single values for their masses, ranging from $10^{-5}$
to 1 M$_\odot$
($10^{-5},\,10^{-4},\,10^{-3},\,10^{-2},\,10^{-1},\,0.5$ and 1
M$_\odot$). We shall refer to these as ``test masses''.

\subsubsection{Bulge geometry}
The most important contribution to self lensing comes from
stellar bulge lenses and/or stars. As the event rates are
proportional to the square root of the lens-source distance, the bulge
geometry may play an important role.
In \citet{kent89}, the bulge is described as an oblate axisymmetric
ellipsoid, and the luminosity density is given as a function of the
elliptical radius $r_e = \sqrt{x^2+y^2+(z/(1-\epsilon(r_e)))^2}$, where $z$
is the distance to the M31 plane and $\epsilon(r_e)$ is the
ellipticity, which varies as a function of the elliptical radius, $r_e$.
The Kent bulge is quite flattened,
and one may wonder if a less flattened model would result in more
self-lensing events. To check this, we have run the \MC\ for a spherical
bulge ($\epsilon=0$), keeping  the total bulge mass and luminosity fixed.
The expected number of both bulge-disc and disc-bulge events
rise both by about 10\%. On the other hand, in absolute terms,
the more numerous contribution of bulge-bulge events decreases by about 5\%
for a net total increase of $\sim 2\%$.
That is, the substitution of a spherical bulge
for an elliptic one has almost no impact on the
total rate of stellar bulge lensing.
This can be traced to the
fact that M31 is seen nearly edge on, which reduces the impact of
distances   perpendicular to the disk.

\subsubsection{Velocities of lenses and sources} \label{sec:vel}

The relative velocities of lenses and sources strongly influence
the rate of microlensing events.
The choice of the velocities adopted in our reference model,
hereafter called model 1, is inspired
from \citet{widrow03} and \citet{geehan05}. We stress that the bulge velocity dispersion
is sensitive not to the mass of the stellar bulge component which contributes to the self-lensing rate, but to the mass of the entire bulge, which may additionally include unknown lensing populations.
We have tested the effect
of changing the bulge velocity dispersion and the M31 disc
rotation velocity in models 2 to 5.
The velocities of the various M31 components
adopted for each model are displayed in Table \ref{tab:vtable}. The solar rotation velocity
is always taken to be 220 km/s and halo dispersion velocities are
always $1/\sqrt{2}$ times the disc rotation velocities.
All velocity dispersions are assumed isotropic,
with the values given being 1-dimensional.

To get an insight into the model dependence of the \MC\ predictions,
it is useful to split the observed spatial region
into an ``inner'' region where most self-lensing events are
expected, and an ``outer'' region which will be dominated by MACHOs if
they are present. We set the boundary between the two regions at an angular
distance of  $8'$ from the centre of M31.

The effect of changing the velocities  for the models
displayed in Table \ref{tab:vtable} is shown in Table \ref{tab:mods}. This
gives the relative change with respect to our reference model
(for a MACHO mass of 0.5 M$_\odot$ and $a_{M31}= 3 \,{\mathrm{kpc}}$).

Beside these normalisation changes,
the distributions of the number of events, as a function of $t_{1/2}$,
the angular distance to the centre of M31, and the maximum
flux increase, all turn out to be almost independent of the model.

\begin{table}[htb!]
\bc
\begin{tabular}{c|cc}
Model &\begin{tabular}{c}bulge velocity \\dispersion (km/s)\end{tabular}&
\begin{tabular}{c}disc rotation \\velocity (km/s)\end{tabular}\\
\hline
1 (reference)&120&250\\
2&120&270\\
3&120&230\\
4&140&250\\
5&100&250\\
\hline
\end{tabular}
\caption{Velocities of M31 components (km/s). The bulge rotation velocity
  and disc velocity dispersion are fixed at 40 km/s and 60
  km/s, respectively.
\label{tab:vtable}}

\begin{tabular}{c|cc|cc}
&\multicolumn{2}{c}{Self Lensing}&\multicolumn{2}{c}{MACHOs}\\
\hline
Model&\begin{tabular}{c}Inner\\ region\end{tabular}&\begin{tabular}{c}Outer\\ region\end{tabular}&\begin{tabular}{c}Inner\\ region\end{tabular}&\begin{tabular}{c}Outer\\ region\end{tabular}\\
\hline
2    &0.97 & 0.98 & 1.15  & 1.21\\
3    &0.97 & 0.96 & 0.84  & 0.81\\
4    &1.03 & 1.03 & 0.98  & 1.01\\
5    &0.92 & 0.90 & 0.98  & 0.99\\
\hline
\end{tabular}
\caption{The velocity dependence of the number of expected events.
The numbers are the ratio
of the number of expected events for models of Table \ref{tab:vtable} to
the same number in the reference model (with M=0.5 M$_\odot$ and
$a_{M31}= 3 \,{\mathrm{kpc}}$). The number of events expected in the reference
model, corrected for detection efficiency, are displayed in Table
\ref{tab:nev-res} of Sect. \ref{sec:results}.
\label{tab:mods}
}
\ec
\end{table}

\subsubsection{Consistency check}
To check the consistency of our \MC, we have computed the optical
depths of the halo both analytically and with the \MC. The results are
identical and consistent with published results \citep{gyuk00,baltz00}.

\subsection{Modelling the observations and the analysis}

The \MC\ generates and selects light curves including part
of the real observational conditions and of the selection algorithm.

Reproducing the photometry conditions in the \MC\ is an
important issue, so we use the same filter as in the real
experiment. This is also true for the colour equations, which relate
fluxes to standard magnitudes in the reference image. In generating
the light curves, all photometric coefficients relating the
observing conditions of the current image to those of the
reference are used  in the Monte Carlo, except for those
related to the seeing correction.

The observation epochs and exposure times reproduce the real ones,
with one composite image per night. In order to avoid
counting the noise twice, no noise has been added to
the \MC\ light curves; it only enters via the error bars.
As we further discuss in Sect. \ref{sec:eff},
an important condition for the efficiency correction to be
reliable is that the \MC\ should not reject events
that the real analysis would have accepted. For this reason, the error
bars in the \MC\ light curves only include the photon noise, and, for
an event to be considered detected, we demand  only the minimum condition that
the corresponding bump rise above the noise
(that is, $L>0$, where
$L$ is the estimator introduced in equation \ref{likelihood}).

\subsection{Event properties}

\begin{figure}[h]
\bc
\includegraphics[scale=0.30]{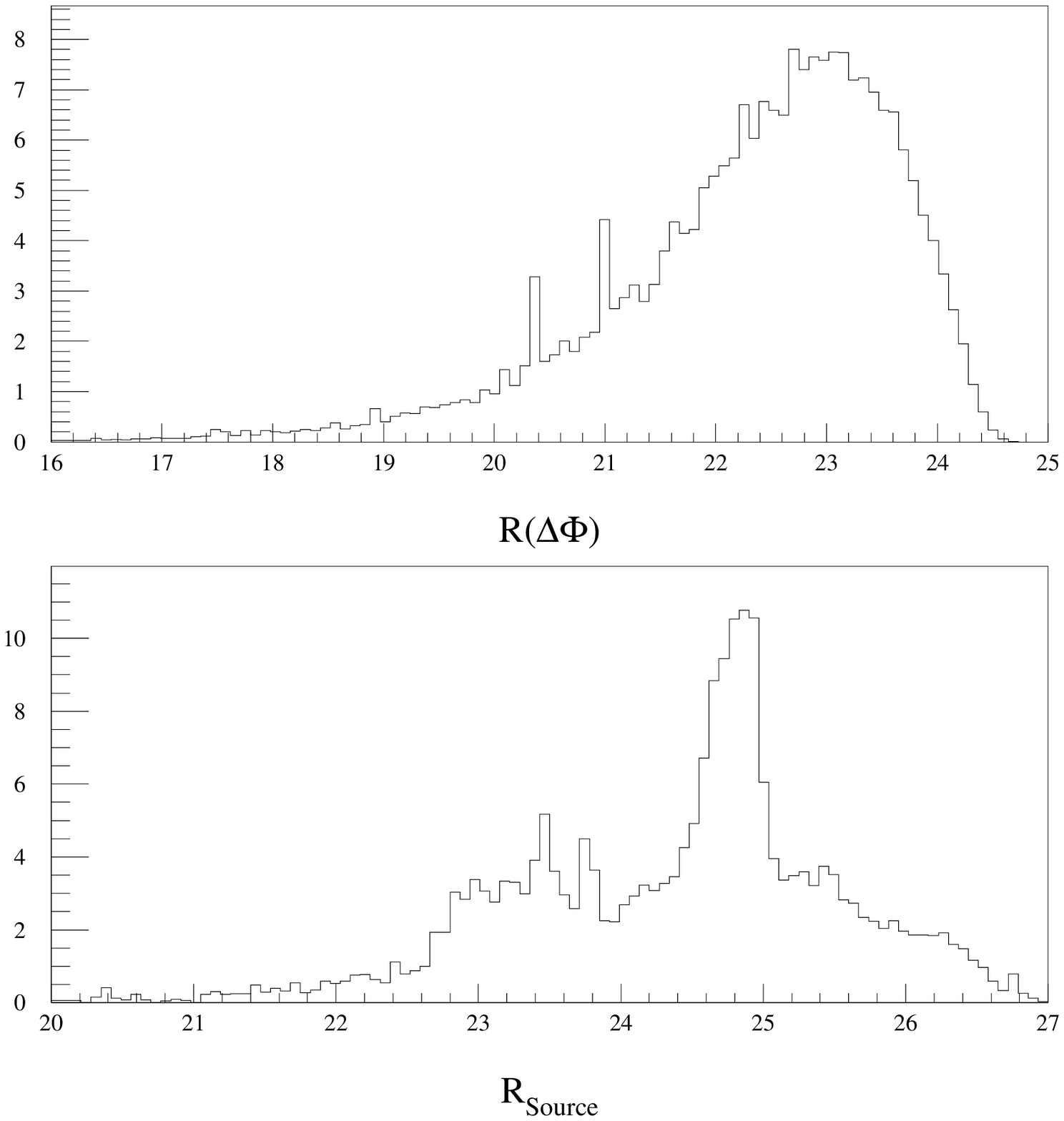}
\includegraphics[scale=0.30]{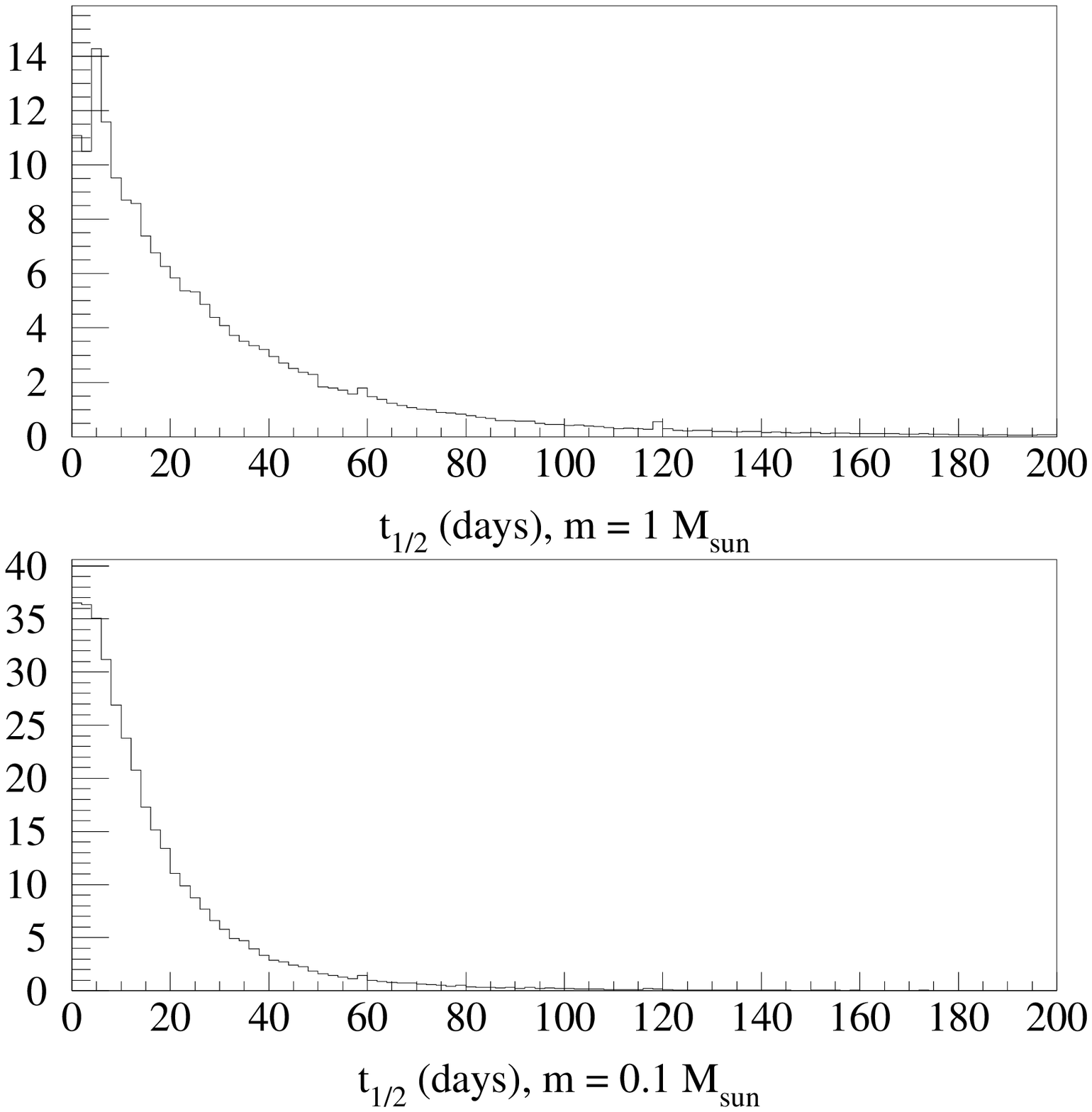}
\ec
\caption{The expected distribution of the R magnitude of the flux increase
  and the source stars (upper panels). The
expected  $t_{1/2}$ distribution for MACHO masses of 1
M$_\odot$ and 0.1 M$_\odot$ (lower panels).
  \label{fig:magnitude}}
\end{figure}

The main observational properties of the events are the R magnitude corresponding to
their flux increase ($R(\Delta\Phi)$) and their duration, which we characterise by
the full-width-at-half-maximum of the bump, $t_{1/2}$.
The CMDs are displayed in Figure
\ref{fig:CM}. We show in Figure \ref{fig:magnitude} the expected
distribution of $R(\Delta\Phi)$, the $R$ magnitude of the sources and
the expected $t_{1/2}$ distribution for two MACHO masses.
The distribution of $t_{1/2}$, quite concentrated toward short durations,
has motivated our choice for the low-duration cutoff
in the selection.

\section{Detection efficiency} \label{sec:eff}

\subsection{The event simulation} \label{sec:eff-sim}

The \MC\  described in the previous section does not
take into account
all the effects we face in the real data analysis.
Therefore, its results, in particular the prediction on the expected number
of events, can only be looked upon as an upper limit.
In order to make a meaningful comparison with the
6 detected events, we must sift the \MC\  results
through an additional filter. This is the
``detection efficiency'' analysis described in this section,
wherein we insert the microlensing
events predicted by the \MC\  into the stream
of images that constitute our actual data set\footnote{We refer to
  this analysis as ``event simulation'', not to be confused with the \MC\
  simulation described in the previous section.}.
This allows us to calculate the detection efficiency
relative to the \MC\  and to obtain
a correct estimate of the characteristics
and total number of the expected events.

The main weakness of the \MC\ in reproducing
the real observations and  analysis stems from the fact that it only
generates microlensing \emph{light curves}, so that it cannot take
into account any aspect related to  \emph{image} analysis.

The \MC\  does not model the background of variable stars,
which both gives rise to high flux variations that can mimic
(and disturb the detection of) real microlensing events,
and generates, from the superposition of many small-amplitude variables,
a non-Gaussian noise that is very difficult to model.

As regards the selection pipeline itself, the \MC\
cannot reproduce the first, essential, cluster detection step
described in Sect. \ref{sec:ana-ml}. Therefore, it cannot test
to what extent the presence in the images
of variations due to the background of variable stars, seeing
variations, and noise, affect the efficiency of cluster detection,
localisation, and separation.

The \MC\  includes neither the seeing
variations nor their correction nor the residuals of the seeing
stabilisation, which also give rise to a non-Gaussian noise.

In principle, it would be possible to reproduce,
within the \MC, the full shape analysis along the light curve
followed in our pipeline. However, the results on the real data turn
out quite different, mainly because the real noise cannot be
correctly modelled analytically.

In practice, no noise is included in the \MC\ light curves,
because the full noise is already present in the images.
Moreover, we have to be careful not to exclude within the \MC\
variations that the real pipeline is able to detect.
As a consequence, the ``shape analysis'' in the \MC\ is quite basic.
We demand only that the  (noiseless) variations reach 3 $\sigma$
above the baseline for three consecutive epochs, where $\sigma$
includes only the photon noise.

The time sampling
of our data set is fully reproduced by the \MC. However, the
sampling criterion along the bump is only implemented in a
very basic way by demanding that the time of maximum magnification lie
within one of the 3 seasons observation.

A typical \MC\  output involves $\sim 20\,000$ events per CCD.
However, adding $20\,000$ events per CCD would significantly alter
the overall statistical properties of the original images (and
therefore of the light curves). In order that the event simulation
provide meaningful results, we cannot add that many events. On the
other hand, the more events we add, the larger the
statistical precision. Particular care has to be taken to avoid
as much as possible simulating  two events so near each other
that their mutual interaction hinders their detectability. Of
course, these difficulties are worse around the centre of the
galaxy, where the spatial distribution of the events is strongly
peaked. Balancing these considerations, we choose to simulate
$5\,000$ events per CCD. The results thus obtained are compatible,
with much smaller errors, with those we obtain by adding only
$1\,000$ events (in which case the crowding problems mentioned
above are negligible).

Each event generated by the \MC\ is endowed with a ``weight''\footnote{As often in \MC\ simulations,
  a weight is ascribed to each generated event. This
  weight carries part of the information on the probability for the
  event to occur, before and independently of any selection in either
  the \MC\ or the event simulation.},
$w_i$, so when we refer to simulated events,
``number'' always means ``weighted number''. Thus  $n_{sim} = \sum_i
w_i$, with statistical error $\Delta n_{sim} = \sqrt{\sum_i w_i^2}$,
where the sum runs over the full set of simulated events.

Let $n_b\equiv n_s+n_r$ be the number of events we simulate on the images,
where $n_s$ and $n_r$ are respectively
the number of selected and rejected events at the end of
the analysis pipeline.
We define the detection efficiency as
$$\varepsilon \equiv \frac{n_s}{n_b},$$ and the relative statistical error is then
$$\left(\frac{\Delta \varepsilon}{\varepsilon}\right)^2 =
\frac{\left(n_r\,\Delta n_s\right)^2+\left(n_s\,\Delta n_r\right)^2}
{\left(n_b n_s\right)^2}.$$
Once we know $\varepsilon$, we can  determine the actual number
of expected events, $n_{exp} = \varepsilon\, n_{exp}^{MC}$,
where $n_{exp}^{MC}$ is the number expected from the \MC\  alone.

The event simulation is performed on the images
\emph{after} debiasing and flatfielding,
but \emph{before} any other reduction step.
We use the package DAOPHOT within IRAF.
First, starting from a sample of $\sim 200$ resolved
stars per CCD, for each image we evaluate the PSF and
the relative photometry with respect to the reference image.
Then we produce a list of microlensing events, randomly chosen among those
selected within the \MC.
For each event,  using  all the light curve
parameters provided by the \MC\ as input, we add to each image the
flux of the magnified star at its position, convolved  with the PSF of the image
(taking due account of the required geometrical and photometric calibration
with respect to the reference image). We then proceed as in the real
analysis. In particular, after image recalibration,
we run the selection pipeline  described in Sect. \ref{sec:ana-ml}.
In short, the scope of the event simulation is to evaluate how many
``real'' microlensing events are going to be rejected by our selection
pipeline.
We test the event simulation procedure  by comparing
the mean photometric dispersion in the light curves
of observed resolved stars to those
of simulated, stable,  stars of comparable magnitude.
We find good agreement.

In the selection pipeline, it is essential to use
data taken in at least two passbands in order to reject variable
objects. Indeed, we test achromaticity with a simultaneous
fit in two passbands and, in the last step of the selection,
we test whether a secondary bump is compatible with being the second
bump of a variable signal. Here, using $i$ band data is important
because the main background arises from long-period, red variable
stars.

In the event simulation, we want to evaluate what fraction of the
\MC\ microlensing events survive the selection
pipeline.  For these genuine microlensing events, we expect the use of
two passbands to be less important. In fact,
microlensing events are expected to pass  the achromaticity
test easily. Moreover, because the events we simulate are  short and bright,
the microlensing bump is in general quite  different from any
possible, very often long\footnote{Short-period variable objects have
already been removed since they are easily recognised from their
multiple variations within the data stream.}, secondary bump, and most
simulated events pass the secondary-bump test. Indeed, we have checked
on one CCD that we get the same result for the detection efficiency
whether we use data in both $r$ and $i$ bands or in  $r$ alone.
For this reason, we have carried out the rest of the event simulation
with $r$-band data only.

\begin{table*}[t]
\bc
\begin{tabular}{|c||c|c|c|}
\hline
criterion & $\varepsilon\, (\Delta\Theta<4')$ & $\varepsilon\,(4<\Delta\Theta<8')$ &
$\varepsilon\,(\Delta\Theta>8')$ \\
\hline
\hline
cluster detection ($Q>100$) &  $46.3\pm 4.1$ & $62.7\pm 1.5$  & $76.4\pm 0.4$\\
\hline
\hline
$L_1>40$ and $L_2/L_1<0.5$ & $40.0\pm 4.0$ & $57.9\pm 1.5$ & $72.5\pm 0.4$\\
$\chi^2/\textrm{dof} < 10$ & $35.7\pm 3.8$ & $54.0\pm 1.5$ & $66.7 \pm 0.4$\\
sampling & $17.1\pm 2.9$ & $31.9\pm 1.4$ & $33.7\pm 0.4$\\
$t_{1/2}<25$ days, $R(\Delta\Phi)<21$ & $14.7\pm 2.8$ & $25.2\pm 1.3$ & $28.5\pm 0.4$ \\
variable analysis & $14.7\pm 2.8$ & $25.2\pm 1.3$ & $28.5\pm 0.4$ \\
\hline
\end{tabular}
\caption{Detection efficiency relative to the \MC\ (in percent), for a MACHO mass M$=0.5$ M$_\odot$,
evaluated at each step of the selection pipeline in different ranges of distance
from the centre of M31.}
\label{tab:ana-eff}

\begin{tabular}{|c||c|c|c|}
\hline
MACHO mass (M$_\odot$) & $\varepsilon\, (\Delta\Theta<4')$ & $\varepsilon\,(4<\Delta\Theta<8')$ &
$\varepsilon\,(\Delta\Theta>8')$\\
\hline
\hline
1 & $19.0\pm 3.0$ & $24.2\pm 1.3$ & $29.7\pm 0.4$ \\
$5\cdot 10^{-1}$ & $14.7\pm 2.8$ & $25.2\pm 1.3$ & $28.5\pm 0.4$\\
$10^{-1}$ & $18.8\pm 3.4$ & $22.1\pm 1.3$ & $26.4\pm 0.4$ \\
$10^{-2}$ & $17.0\pm 3.7$ & $21.8\pm 1.6$ & $23.5\pm 0.5$ \\
$10^{-3}$ & $10.1\pm 3.2$ & $14.1\pm 1.6$ & $15.6\pm 0.5$ \\
$10^{-4}$ & $2.4\pm 1.5$ & $8.9\pm 2.5$ & $9.5\pm 0.5$ \\
$10^{-5}$ & $0.37\pm 0.43$ & $5.4\pm 2.2$ & $6.2\pm 0.7$ \\
\hline
\hline
self lensing & $17.8\pm1.2$ & $22.6\pm 0.6$ & $26.9\pm 0.3$\\
\hline
\end{tabular}
\caption{Detection efficiency relative to the \MC\ (in percent),
for our test set of  MACHO masses and for self lensing, for the
same distance ranges as in Table \ref{tab:ana-eff}.}
\label{tab:res-eff}

\ec

\end{table*}

\subsection{The results} \label{sec:eff-res}

For each CCD (with 4 CCDs per field) we simulate
at most 5000 microlensing events, randomly chosen among those selected
within the \MC, and subject to conditions reflecting the selection cuts.
We only simulate events that are both bright
($R(\Delta\Phi)<21.2$) and short ($t_{1/2}<27$ days).
These thresholds are looser than those used in the selection
($R(\Delta\Phi)<21.0$ and $t_{1/2}<25$ days)
because we want to include all events that can in principle be detected
by the pipeline. These enlarged cuts reflect the
dispersion of the difference between the input and output event
parameters of the event simulation.
To test this choice, we have also run some test jobs
using slightly different input cuts. For instance, if one uses the looser cuts
$R(\Delta\Phi)<21.5$ and $t_{1/2}<30$ days, the number of events
predicted by the \MC\ is larger, but the efficiency turns out to be
smaller. The two effects cancel, and the end result for the number of
expected events corrected for detection efficiency remains
unchanged. For each CCD we run the event simulation for our test masses. As in
the real analysis, we mask
the very central region of M31.

The detection
efficiency depends mainly on the distance from the centre of M31,
the time width, and the maximum flux increase.
We  run the event simulation only for model 1
(Sect. \ref{sec:vel}) and a M31 core radius $a_{M31}=5 \,{\mathrm{kpc}}$.
In fact, there is no reason for the efficiency at a given position
in the field to depend
on the core radius. It could in principle depend on
distributions of the time width and the maximum flux increase,
but we have seen that these distributions
are almost model-independent.

Finite-source effects can  produce significant deviations
from a simple \pacz\ shape,
and this can be quite important toward M31,
where most sources are giant stars.
We expect this effect to be particularly relevant
for low mass MACHOs.
The events generated by the \MC\ (Sect. \ref{sec:MC}) and entered in
the event simulation include finite-source effects,
although the microlensing fit in the selection pipeline uses only simple
\pacz\ curves. This  causes an efficiency loss, which we evaluate as
follows: we run an event simulation, for one CCD and all test masses,
without including finite-source effects in the input events, and then
evaluate the associated efficiency rise.
This ought to be of the same order as the efficiency loss in the
real pipeline.
For masses down to 10$^{-2}\,M_\odot$
the change turns out to be negligible. For masses smaller or equal to
10$^{-3}\,M_\odot$, it is of the order of 20\% or less.

The detection efficiency depends on position in the
field primarily through the distance to the centre of M31. At a given distance
we find no significant difference between the various
CCDs. At angular distances larger than 8' the efficiency is practically
constant. In the region inside 8', the efficiency steadily decreases
toward the centre. This can be traced to the increase of both the
crowding and the surface brightness. Indeed, the
drop of  efficiency in the central region mainly comes from the
first step of the selection pipeline, namely the cluster detection.

Table \ref{tab:ana-eff} shows the contribution of the successive
steps of the analysis to the total loss of efficiency.
The distance to the centre of M31 is divided into 3 ranges
($\Delta\Theta<4',\,4'<\Delta\Theta<8'$ and $\Delta\Theta>8'$). The
MACHO mass is $0.5\,M_\odot$ but the qualitative features discussed
below are the same for all masses.
We have isolated the first step of the analysis, the cluster detection,
which is implemented on the images, while the others are performed on
the light curves. As emphasised earlier, the
increase in crowding and surface brightness near the centre
causes a significant drop of efficiency in the two central regions.
Most of the dependence of the efficiency on the distance
to the centre arises from this step, whereas the effects of all other steps,
acting on light curves, are nearly
position independent. Note the loss of
efficiency by almost a factor of 2 associated with the sampling
cut. This is not surprising as this cut is implemented in the
\MC\ in only a very basic way.

Table \ref{tab:res-eff} gives
the detection efficiency for our test set of MACHO masses after the full event selection.
Down to a mass of $10^{-2}$ M$_\odot$,
we find no significant differences
between self-lensing and MACHO events.
This reflects the fact that their main
characteristics do not differ significantly  on average.
For very small masses, we find
a drop in the efficiency, due to both the smaller time widths of the
bump and finite-source effects.

\section{Results and halo fraction constraints} \label{sec:results}

In this section, we present the result of the complete simulation, the
\MC\ followed by the event simulation, and
discuss what we can infer about the fraction $f$ of MACHOs present in the
halos of M31 and the MW from the comparison with the data presented in Sect.
\ref{sec:events}.

\begin{figure}[hbt!]
\resizebox{\hsize}{!}{\includegraphics{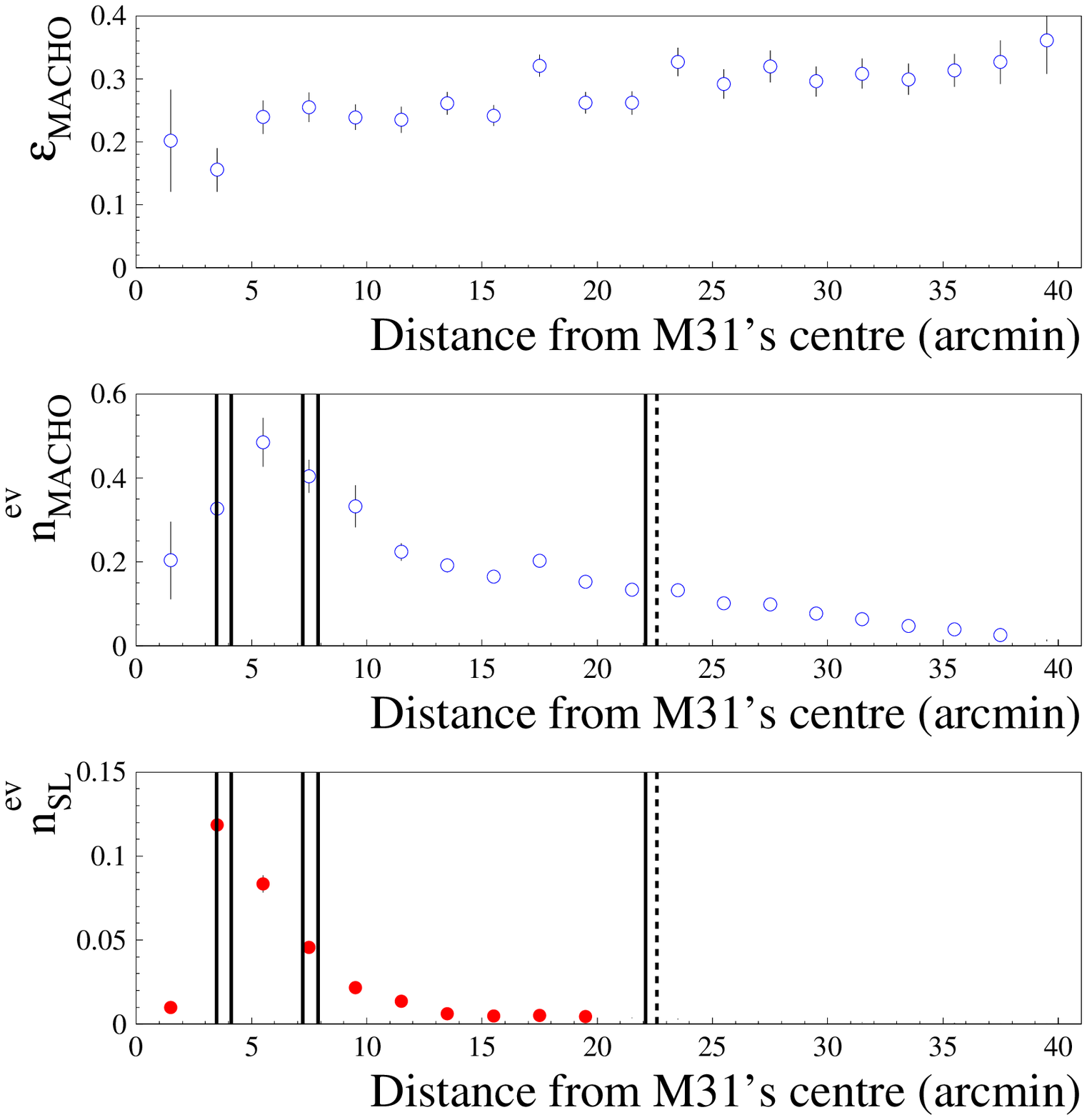}}
\caption{
Predictions of the full simulation as a function of the distance from
the centre of M31. Upper pannel: efficiency correction (for MACHOs);
central panel: expected number of MACHO events (full halo, $M=0.5$
M$_\odot$, $a_{M31}= 3\,{\mathrm{kpc}}$); bottom panel: expected number of
self-lensing events (for a stellar bulge (disc) $M/L_B=3\,(4)$).
The vertical lines indicate the position of the observed events, the
dashed line corresponds to PA-00-S4, which has been excluded from the analysis
because it is a probably M31/M32 intergalactic event.}
\label{fig:sl-dark-vs-d}.
\end{figure}

\begin{table*}[tbh]
\bc
\begin{tabular}{|c||c|c|c|c|}
\hline
 & $\Delta\Theta<4'$ & $4<\Delta\Theta<8'$ &
$\Delta\Theta>8'$\\
\hline
\hline
mass (M$_\odot$) \\
\hline
Halo, $a_{M31}$ = 3 kpc&&& \\
1                & $0.70\pm 0.12$   & $1.38\pm 0.08$   & $2.96\pm 0.04$  \\
$5\cdot 10^{-1}$ & $0.81\pm 0.17$   & $1.93\pm 0.11$   & $4.18\pm 0.08$\\
$10^{-1}$        & $1.63\pm 0.32$   & $3.00\pm 0.20$   & $8.10\pm 0.23$  \\
$10^{-2}$        & $1.93\pm 0.45$   & $3.85\pm 0.30$   & $12.65\pm 0.29$  \\
$10^{-3}$        & $0.72\pm 0.27$   & $2.20\pm 0.30$   & $9.17\pm 0.29$\\
$10^{-4}$        & $0.064\pm 0.042$ & $0.60\pm 0.18$   & $3.09\pm 0.18$ \\
$10^{-5}$        & $0.002\pm 0.002$ & $0.034\pm 0.015$ & $0.42\pm 0.06$\\
\hline
\hline
Halo, $a_{M31}$ = 5 kpc&&&\\
1                & $0.60\pm 0.10$   & $1.11\pm 0.06$   & $2.48\pm 0.04$  \\
$5\cdot 10^{-1}$ & $0.74\pm 0.18$   & $1.57\pm 0.09$   & $3.63\pm 0.09$\\
$10^{-1}$        & $1.30\pm 0.25$   & $2.52\pm 0.16$   & $6.94\pm 0.12$  \\
$10^{-2}$        & $1.41\pm 0.34$   & $3.63\pm 0.29$   & $11.29\pm 0.24$  \\
$10^{-3}$        & $0.81\pm 0.30$   & $2.07\pm 0.26$   & $8.41\pm 0.26$\\
$10^{-4}$        & $0.15\pm 0.15$   & $0.49\pm 0.15$   & $2.83\pm 0.16$ \\
$10^{-5}$        & $0.002\pm 0.002$ & $0.048\pm 0.022$ & $0.40\pm 0.05$\\
\hline
\hline
self lensing   & $0.29\pm 0.02$  & $0.29\pm 0.01$ & $0.16\pm 0.01$\\
\hline
\end{tabular}
\caption{The expected number
of MACHO and of the self-lensing  events, corrected for  efficiency,
for the models with $a_{M31}=3\,{\mathrm{kpc}}$ and
$a_{M31}=5 \,{\mathrm{kpc}}$, in three different ranges of distance from the M31
centre. The stellar bulge (disc)  $M/L_B$ ratio is equal to $3\,(4)$.
}
\label{tab:nev-res}
\ec
\end{table*}

In Table \ref{tab:nev-res} we present the expected numbers of
self-lensing and halo events (for a full halo and two different
values of the core radius) predicted by the full simulation in
the three distance ranges $\Delta\Theta<4',\,4<\Delta\Theta<8'$
and $\Delta\Theta>8'$. The self-lensing results, given for a stellar bulge
$M/L_B$ ratio equal to 3, are dominated by stellar bulge lenses and
therefore scale with this ratio. This must be compared with the 5
microlensing events reported in Sect. \ref{sec:events}. PA-00-S4,
which is located near the line of sight toward the M32 galaxy, is
likely an intergalactic microlensing event \citep{paulin02} and
therefore not included in the present discussion. Accordingly,
we have excluded from the analysis a 4' radius circular region
centred on M32.

The main issue we have to face is distinguishing
self-lensing events from halo events. This is particularly
important  as the number of expected MACHO and self-lensing
events is of about the same order of magnitude
if the halo fraction is of order 20\% or less as in the
direction of the Magellanic clouds.

\begin{table*}[tbh]
\bc
\begin{tabular}{c|ccc||ccc}
&\multicolumn{3}{c}{$a_{M31}=3$ kpc}&\multicolumn{3}{c}{$a_{M31}=5$ kpc}\\
\hline
Mass (M$_\odot$)&$f_\mathrm{INF}$&$f_\mathrm{MAX}$&$f_\mathrm{SUP}$&$f_\mathrm{INF}$&$f_\mathrm{MAX}$&$f_\mathrm{SUP}$\\
\hline
1                   & 0.27& 0.81& 0.97 & 0.29& 0.97& 0.97 \\
$5\cdot 10^{-1}$    & 0.22& 0.57& 0.94 & 0.24& 0.67& 0.96 \\
$10^{-1}$           & 0.13& 0.31& 0.74 & 0.15& 0.37& 0.83 \\
$10^{-2}$           & 0.08& 0.21& 0.51 & 0.09& 0.23& 0.57 \\
$10^{-3}$           & 0.11& 0.29& 0.73 & 0.12& 0.31& 0.76 \\
$10^{-4}$           & 0.20& 0.77& 0.96 & 0.18& 0.81& 0.96 \\
$10^{-5}$           & 0.12& 1.00& 0.97 & 0.10& 1.00& 0.97 \\
\hline
\end{tabular}
\caption{Results for the halo fraction $f$: the 95\% CL lower bound
($f_\mathrm{INF}$) and upper bound ($f_\mathrm{SUP}$), and
maximum probability ($f_\mathrm{MAX}$) are displayed for
 $a_{M31}=3 \,{\mathrm{kpc}}$ and $a_{M31}=5\,{\mathrm{kpc}}$. In both cases,
the stellar bulge (disc) $M/L_B$ ratio is $3\,(4)$.
\label{tab:frac-res-l1}
}
\ec
\end{table*}

Although the observed characteristics of the light curves
do not allow one to disentangle  the two classes of events,
the spatial distribution of the detected events (Fig. \ref{fig:field})
can give us useful insights.
While most self-lensing events are
expected in the central region, halo events should be more evenly
distributed out to larger radii. In Figure \ref{fig:sl-dark-vs-d},
together with the distance dependence of the detection efficiency, we show
the expected spatial distribution of self lensing and 0.5 M$_\odot$
MACHO events (full halo). The observed events
are clustered in the central region with the significant
exception of PA-99-N2, which is located in a region where the self-lensing
contamination to MACHOs events is expected to be small.

The key aspect of our analysis is the comparison
of the expected spatial distribution of  the events
with that of the observed ones.
In order to carry out this comparison as precisely as possible,
we divide the observed field into a large number
of bins, equally spaced in distance from M31's centre.
We present here an analysis with 20 bins of 2' width, but we have checked
that the results do not change significantly if we use either 40 bins of 1' width
or 10 bins of 4' width.

\subsection{The halo fraction} \label{sec:halof}

The first striking feature in the comparison between predictions and
data  is that we observe far more events than predicted
for self lensing alone.
Therefore, it is tempting to conclude that the events in excess
with respect to the prediction should be considered as MACHOs.
This statement can be made more quantitative:
given a MACHO halo fraction, $f$, we can compute
the probability of getting the observed number of events
and, by Bayesian inversion,  obtain the probability
distribution of the halo fraction.

As already outlined, we bin the
observed space into $N_{bin}$ equally spaced annuli
and then, given the model predictions $x_i$ ($i=1\ldots N_{bin}$),
obtain the combined probability
of observing in each bin $n_i$  events.
The combined probability is  the product of the individual probabilities
of independent variates $n_i$:
\begin{equation}\label{eq:prob}
P\left(n_i|x_i\right)=
\prod^{N_{bin}}_{i=1}
\frac{1}{n_i!}
\exp(-x_i) x_i^{n_i}.
\end{equation}

For a given a model, the different $x_i$ are not independent:
they all depend on the halo fraction $f$ via the  equations
\begin{equation}\label{eq:model}
x_i  =  h_i\,f+s_i,
\end{equation}
where $h_{i}$ and  $s_{i}$ are the numbers of events
predicted in bin $i$ for a full MACHO halo
and self lensing, respectively. A model specifies $h$ and $s$,
so the probability depends on only one parameter, $f$.
It is therefore possible to evaluate lower and upper limits
at a given confidence level for the halo fraction $f$.

\begin{figure}
\resizebox{\hsize}{!}{\includegraphics{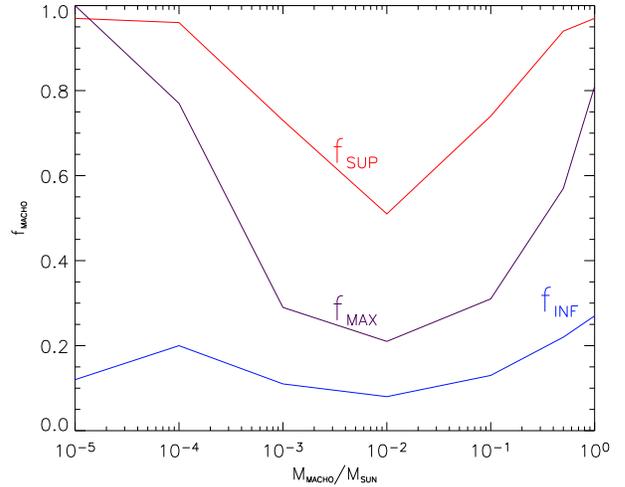}}
\caption{
Most probable value,  upper and lower 95\% CL limit for the
halo fraction as a function of the MACHO mass
for $a_{M31}=3\,{\mathrm{kpc}}$ and stellar bulge (disc) $M/L_B=3\,(4)$.}
\label{fig:halof1}
\end{figure}

In Figure \ref{fig:halof1} and Table \ref{tab:frac-res-l1}, we display the
95\% confidence level (CL) limits obtained in this configuration for $a_{M31}=3 \,{\mathrm{kpc}}$
and $M/L_B=3$. We get a significant lower limit, $f_\mathrm{INF}> 20\%$, in the
mass range from 0.5 to 1 $M_\odot$. No interesting upper bound on $f$ is
obtained except around a mass of $10^{-2}\,M_\odot$ ($f_\mathrm{SUP}=50\%$).
We also show in Table \ref{tab:frac-res-l1} the same limits for
$a_{M31}=5\,\mathrm{kpc}$. As the predicted halo contribution is
smaller, the inferred lower limit on $f$ is slightly larger.

\subsection{Self-lensing background ?} \label{sec:sl}

The fact that 4 out of the 5 observed events lie within 8' from
the centre of M31 could be suggestive of
self-lensing origin, implying that we underestimate this
contribution. However,  in the \MC\ section we have
already seen that the velocity dependence of our results is very
weak. For models 2 (3), where the change is maximum, the 95\% CL
lower limit on $f$ in the mass range 0.1-1 M$_\odot$ is shifted
by about - (+) 0.02.
Furthermore, $M/L_B$ ratios larger
than 4 cannot be accommodated by known stellar populations.
Still, for comparison, we have considered models for which, on dynamical grounds,
the $M/L_B$ ratio of either the disc or the bulge take values up to $\sim 8-9$.
One can see from Table \ref{tab:mldyn} that our conclusions are not qualitatively
altered. This can be partly attributed to the occurence of
PA-99-N2 $22'$ away from the M31 centre.

\begin{table}[htb]
\bc
\begin{tabular}{c|c||c|c||c}
Bulge ${M/L}_B$&Disc ${M/L}_B$ & $n_{SL}$ & $P(f=0)$ & $f_\mathrm{INF}$\\
\hline
3 & 4 & 0.72 & $10^{-4}$&0.22\\
3 & 9 & 1.1 & $10^{-3}$ &0.17\\
8 & 4 & 1.5 & $4\,10^{-3}$&0.15\\
\hline
\end{tabular}
\caption{For different sets of values of stellar bulge and disc
$M/L_B$ (Sect. \ref{sec:lenses}) we report the number of expected self-lensing events, corrected
for the efficiency, the probability for the signal to be a Poisson fluctuation
for a $f=0$ halo  and, for a $M=0.5\, \mathrm{M}_\odot$ MACHO population MW and M31
halos with $a_{M31}=3\,\mathrm{kpc}$,
the 95\% CL lower bound for the halo fraction $f$.
\label{tab:mldyn}
}
\ec
\end{table}

One can also question the
bulge geometry. However, we have seen that assuming a spherical bulge with
the same mass and luminosity does not alter the results. One could
also think of a bar-like bulge.
This possibility has been considered by \citet{gerhard:86}, who has shown that unless a would be
 bar points toward us within $10^\circ$, its ellipticity does
not exceed 0.3. This cannot produce a
significant increase  of the self-lensing prediction.
Even if a bar-like bulge points toward us and is highly prolate, it
cannot explain event PA-99-N2.

Clearly, unless we grossly misunderstand the bulge of
M31, our events cannot be explained by self lensing alone.

Still, in view of our low statistics, we could be
facing a Poisson fluctuation. However, this is highly improbable: given the prediction of our
simulation, the probability of observing 5 self-lensing
events with the observed spatial distribution is $P(f=0) \sim 10^{-4}$
for a $M/L_B =  3\,(4)$ M31 stellar bulge (disc), and remains well below $\sim 10^{-2}$
even for much heavier configurations (Table \ref{tab:mldyn}).

\section{Conclusions} \label{sec:conclusion}
In this paper, we present
first constraints on the halo fraction, $f$, in the form
of MACHOs in the combined halos of M31 and MW, based
on a three-year search for gravitational microlensing  in the
direction of M31.

Our selection pipeline, restricted to
bright, short-duration variations, leads us to the detection
of 6  candidate microlensing events. However, one of these
is likely to be a M31-M32 intergalactic self-lensing event, so
we do not include it when assessing the halo fraction $f$.

We have thoroughly discussed the issue of the possible contamination
of this sample by background variable stars.
Indeed, we are not aware of any class of variable stars able
to reproduce such light curves, therefore we have assumed
that  all our candidates are \emph{genuine} microlensing events.

To be able to draw physical conclusions from this result,
we have constructed a full simulation of the expected results, which
involves a \MC\  simulation completed by an event simulation to account
for aspects of the observation and the selection pipeline not included
in the \MC.

The full simulation predicts that M31 self lensing alone should give
us less than 1 event,  whereas we observe 5, one of which
is located $22'$ away from the M31's centre, where the expected
self-lensing signal is negligible. As the probability that we
are facing a mere Poisson fluctuation from the self-lensing
prediction is very small ($\sim 0.01\%$), we consider these results
as  evidence for the detection of
MACHOs in the direction of M31. In particular, for $a_{M31}=3\,\mathrm{kpc}$ and
a $M/L_B$ ratio for the disc and stellar bulge smaller than 4,
we get a 95\% CL lower limit of $20-25\%$ for $f$, if the average
mass of MACHOs lies in the range 0.5-1 $M_\odot$. Our signal is compatible
with the one detected in the direction of the Magellanic clouds by
the MACHO collaboration \citep{macho00}.

We have also considered models that, on dynamical grounds, involve
higher disc or stellar bulge $M/L_B$ ratios.  However, because of the spatial
distribution of the observed events, the
conclusion would not be qualitatively different. Indeed, because
of the presence of the event PA-99-N2 $22'$ away from the M31 centre
where self lensing is negligible, the lower bound  on $f$
would not pass below $\sim 15\%$ even in the most extreme models considered.

Finally, the observed events can hardly be blamed on
the geometry of the bulge. Indeed, the number of predicted
self-lensing events cannot be significantly increased unless it has a highly prolate
bar-like structure exactly pointing toward us. However, even this improbable
configuration would not explain one of the events, which definitely
occurs outside the bulge.

Beside the 5 events selected by our pipeline, we have found
a very likely candidate for a binary lensing event with caustic
crossing.
This event occurs $\sim 32'$
away from M31's centre, where one can safely ignore  self
lensing. Therefore, although included in neither our selection pipeline nor
our discussion on the halo fraction,
this detection strengthens our conclusion that we are detecting  a MACHO
signal in the direction of M31.

To get more stringent constraints on the
modelling of M31, better statistics are badly needed.
To achieve this goal using our data,  we  plan
to extend the present analysis in a forthcoming work by  looking  for
fainter variations.
Another option would be to lift the duration cut.
However, we consider this less attractive,
because the contamination by the background of variable stars would be
much larger and difficult to eliminate.
Moreover, the \MC\ predictions
disfavour a major contribution of long duration events.

\paragraph*{\emph{Note added in proof.}}~~
After submission of this work, the MEGA collaboration presented their
results obtained independently from the same data (De Jong et al.,
[arXiv:astro-ph/0507286 v2]). Their conclusions are different from ours.
We would like to point out that their criticism  of our analysis
is not relevant because, as stated in Section 4.1.2, we choose to only
consider for self lensing evaluation a population of stars with a
standard M/L ratio, which does not need to account for the total dynamical mass nor
to reproduce the inner rotation curve.

\begin{acknowledgements}
SCN was supported by the Swiss National Science Foundation.
JA was supported by a Leverhulme grant.
AG was supported by grant AST 02-01266 from the US NSF.
EK was supported by an Advanced Fellowship from the Particle
Physics and Astronomy Research Council (PPARC).
CSS was supported by the Indo French Center for Advanced Research (IFCPAR)
under project no. 2404-3.
YT was supported by a Leverhulme grant.
MJW was supported by a PPARC studentship.
\end{acknowledgements}
\bibliographystyle{aa}
\bibliography{bibli2}
\end{document}